\documentstyle[psfig]{mn}

\def\etal{ et~al. }
\def\micron{\hbox{$\mu\rm m$}}

\title[The AAO's 2dF Facility]
 {The Anglo-Australian Observatory's 2dF Facility}
\author[Lewis \etal]{I. J. Lewis,$^1$\thanks{Correspondence address: Department of Physics, Keble Road,
	  Oxford OX1 3RH, UK. E-mail: ijl@astro.ox.ac.uk}
       R. D. Cannon,$^1$
       K. Taylor,$^{1,2}$
       K. Glazebrook,$^{1,3}$
       J. A. Bailey,$^1$\cr
       I. K. Baldry,$^{1,3}$
       J. R. Barton,$^1$
       T. J. Bridges,$^1$
       G. B. Dalton,$^4$
       T. J. Farrell,$^1$\cr
       P. M. Gray,$^{1,5}$
       A. Lankshear,$^1$
       C. McCowage,$^1$
       I. R. Parry,$^{1,6}$
       R. M. Sharples,$^7$\cr
       K. Shortridge,$^1$
       G. A. Smith,$^1$
       J. Stevenson,$^1$
       J. O. Straede,$^1$
       L. G. Waller,$^1$\cr
       J. D. Whittard,$^1$
       J. K. Wilcox$^1$
       and K. C. Willis$^{1,8}$
              \\
  $^1$Anglo-Australian Observatory, PO Box 296, Epping, NSW 1710, Australia \\
  $^2$Department of Astronomy, California Institute of Technology, Pasadena
  California 91125-2400, USA \\
  $^3$Department of Physics and Astronomy, Johns Hopkins University, 3400 North Charles Street
  Baltimore, MD 21218-2686, USA \\
  $^4$Department of Physics, Keble Road, Oxford OX1 3RH, UK \\
  $^5$ESO Paranal Observatory, PO Box 540, Antofagasta, Chile \\
  $^6$Institute of Astronomy, Madingley Road, Cambridge CB3 0HA, UK \\
  $^7$Department of Physics, University of Durham, South Road, Durham DH1 3LE, UK \\
  $^8$Australian Centre for Field Robotics, Rose Street, Building J04, University of Sydney,
  NSW 2006, Australia\\
}

\date{\today}

\pagerange{\pageref{firstpage}--\pageref{lastpage}}
\pubyear{2001}

\begin{document}

\maketitle

\label{firstpage}

\begin{abstract}

The 2dF (Two-degree Field) facility at the prime focus of the
Anglo-Australian Telescope provides multiple object spectroscopy over
a $2^\circ$ field of view.  Up to 400 target fibres can be
independently positioned by a complex robot.  Two spectrographs
provide spectra with resolutions of between 500 and 2000, over
wavelength ranges of 440nm and 110nm respectively.  The 2dF
facility  began routine observations in 1997.

2dF was designed primarily for galaxy redshift surveys and has a
number of innovative features. The large corrector lens incorporates
an atmospheric dispersion compensator, essential for wide wavelength
coverage with small diameter fibres.  The instrument has two full sets
of fibres on separate field plates, so that re-configuring can be done
in parallel with observing.  The robot positioner places one
fibre every 6~seconds, to a precision of 0.3~arcsec (20~\micron) over
the full field.  All components of 2dF, including the
spectrographs, are mounted on a 5-m diameter telescope top-end ring
for ease of handling and to keep the optical fibres short in order to
maximise UV throughput.

There
is a pipeline data reduction system which allows each data set to be fully
analysed while the next field is being observed.

2dF has achieved its initial astronomical goals.  The
redshift surveys obtain spectra at the rate of 2500
galaxies per night, yielding a total of about 200,000 objects in the
first four years.  Typically a B$=19$ galaxy gives a spectrum
with signal to noise ratio of better than 10 per pixel in less than an hour; redshifts are
derived for about 95 per cent of all galaxies, with 99 per cent reliability or
better.  Total system throughput is about 5 per cent.  The failure rate of
the positioner and fibre system is about 1:10,000 moves or once every
few nights and recovery time is usually short.

In this paper we provide the historical background to the 2dF
facility, the design philosophy, a full technical description
and a summary of the performance of the instrument.  We also briefly
review its scientific applications and possible future developments.

\end{abstract}

\begin{keywords}
instrumentation: spectrographs -- techniques: spectroscopic -- surveys
-- galaxies: distances and redshifts -- large-scale structure of Universe
\end{keywords}

\section{Introduction}
\label{intro}

The value of equipping telescopes with a large field of view has been
recognised for some time. In 1986 the Royal Astronomical Society
report ``Review of Scientific Priorities for UK Astronomical Research
1990--2000'' \cite{ras} put a wide-field multi-object spectroscopic
facility at the top of its priority list for new projects.
Subsequently the UK Large Telescope Panel recommended that a
wide-field survey facility be pursued in tandem with an 8 metre
telescope project.

The general scientific case for a wide-field spectroscopic facility on
a 4m telescope was two-fold: to provide spectra for large samples of
objects found in the multi-colour imaging surveys from the UK, ESO and
Oschin (Palomar) Schmidt telescopes; and to generate targets for the
coming generation of 8-10m optical telescopes.  The biggest specific
science driver was to obtain redshifts of tens or even hundreds of
thousands of galaxies and quasars, to elucidate the three-dimensional
structure and evolution of the universe.  Other major projects
required spectra for large samples of stars, to determine their
kinematics and composition and hence the dynamical and chemical
history of our Galaxy; for similar studies of the Magellanic Clouds;
and for detailed studies of star clusters and clusters of galaxies.

One possibility was to provide a multi-fibre upgrade to the 3.9 metre
Anglo-Australian Telescope (AAT). This was particularly advantageous
for two reasons. Firstly, the optical design of the telescope (a
hyperboloidal primary mirror and relatively slow f-ratio of f/3.3 at
prime focus) enabled a wide field of 2\degr\ to be achieved using a
large but straightforward optical corrector. Secondly, the
Anglo-Australian Observatory (AAO) already had extensive experience
with multi-object fibre spectroscopy. This dates from the pioneering
days of optical fibres and its brass plug-plate system FOCAP
\cite{focap} and more recently the fully automated AUTOFIB fibre
positioner system \cite{autofib}.

At the end of 1988 the AAT Board (AATB) commissioned a full design
study of a wide-field fibre-optic spectroscopic facility for the AAT
\cite{designsummary}. Following
further detailed investigations into cost and budgets, and in the
expectation of some additional funding from both the Australian and UK
Governments, the AATB gave its approval to begin the Two-degree Field
(or 2dF) project in March 1990.  Initially the direct budget allocated
was A\$2.25M (for components and certain specific tasks which could
not be done in-house) and it was expected to take 4-5
years to complete.  Since the project would dominate the AAO's
activities for several years, a Project Management Committee was
established with several external expert members.

In developing the 2dF facility the AAO wished to build on the techniques
already used for the fully automated AUTOFIB instrument in use at the
Cassegrain focus of the AAT. Since 2dF would offer almost an order of
magnitude increase in multiplex advantage over AUTOFIB and would be
located at the more challenging prime focus, several
technical problems had to be solved before final approval was given
for commencement of the project.

The technical problems were addressed in the initial design study
reports and covered areas such as the accuracy required of the robotic
positioner, the design of the fibre retraction systems, location and
design of the fibre spectrographs and the requirement for a double
fieldplate system to maximise observing time.

The mechanical constraints were eased by the strength and rigidity of
the telescope tube structure and the size of the dome, which meant that
a large instrument could be housed at prime focus without clearance
and flexure problems.

Commissioning of the 2dF facility began with the new prime focus
corrector in the latter part of 1993. The instrument was officially
declared open at a ceremony on 1995 November 20, and the first
spectroscopic data were obtained in mid-1996. The facility began to
provide scheduled scientific observations in September 1997 with
almost full functionality.  The project ran about 40 per cent over
the original time estimate and 20 per cent over budget (these two are linked,
in that 2dF could have been completed sooner had more funds been
available).   The
effective total cost was subsequently estimated to have been about
A\$8M, including all staff costs and overheads.  The bulk of the
design and construction work was eventually done in-house using the
AAO's facilities in Sydney and Coonabarabran, partly because many
aspects involved innovative design features which could not be easily
specified or contracted out, and partly to contain costs.

Progress reports describing the evolution of the design of 2dF have
been published
\cite{progress1,progress2,progress3,progress4,progress5,progress6} and
project updates have featured regularly in the quarterly AAO Newsletter.

In this paper we provide a full technical description of the 2dF
facility and its performance. Section \ref{overview} gives an overview
of the instrumentation and its relationship to other multiple object
fibre spectroscopy systems. In section \ref{components}, a detailed
description of the individual components of the 2dF facility is
given. Operation of the 2dF facility is covered in section
\ref{operations} and actual performance is detailed in section
\ref{performance}. Sections \ref{science} and \ref{future}
describe the range of projects being done with
2dF and some scientific possibilities for the future.
Up to date technical information and signal to
noise calculators are available on the 2dF WWW pages
(http://www.aao.gov.au/2df/).

\section{Overview of instrument}
\label{overview}

\subsection{Objectives and design philosophy}

The 2dF facility was developed with the aim of providing the AAT with
a dedicated prime focus spectroscopy facility, with order of magnitude
improvements over existing systems in terms of the field area and
number of objects that it is possible to observe simultaneously (the
multiplex advantage).  While 2dF was designed from the outset to be a
versatile common-user instrument, suitable for a wide range of
astronomical projects on an international research facility, it was
always clear that the dominant project would be to obtain redshifts
and hence distances for a very large sample of galaxies (a few times
$10^{5}$ objects), to map out the three-dimensional structure of the
relatively local universe.

Two separate but linked large redshift surveys where the initial main projects
with 2dF:
one for 250,000 galaxies brighter than b$_J$=19.45 with a median redshift of 0.1
\cite{grs} and one for 30,000 colour-selected quasars covering redshifts up to 4
\cite{qrs}. This meant that 2dF had to be optimised to obtain low dispersion
spectra of a few thousand targets per night, over a wide wavelength
range, for mainly non-stellar objects which were in the magnitude range $15 < B <
20$  and spread across more than 1000 square degrees of sky.  These
requirements are very well met by a system providing two sets of 400
fibres covering 3~square degrees of sky, and capable of
reconfiguring one set of fibres in about an hour, which happens to be
equal to the time needed to take adequate signal-to-noise
ratio  spectra of 19-20 magnitude galaxies.

Whenever choices had to be made on design features, instrument
parameters or priorities, the driver was to maximise 2dF's efficiency
as a redshift engine.  The requirements of other projects, such as
taking higher resolution spectra of stars, doing longer integrations
on fainter galaxies, or working on densely clustered targets, were
incorporated where possible, but not if they involved any compromise
of performance for the redshift surveys.

\subsection{Other multi-fibre systems}

Previous multi-object fibre spectroscopy systems on 4 metre class
telescopes have used smaller fields of view and significantly smaller
multiplex advantage.  For example the William Herschel Telescope's
AUTOFIB-2 \cite{af2} and WYFFOS \cite{wyffos} instruments provide
fibre spectroscopy of up to 150 objects over a 1\degr\ field
(40 arcminutes unvignetted). The HYDRA instrument \cite{hydra} on the 3.5m
WIYN telescope uses approximately 100 optical fibres to cover an
unvignetted field of 1\degr.  Earlier fibre instruments on the AAT
were FOCAP with 50 fibres \cite{focap} and AUTOFIB with 64 fibres
\cite{autofib}, both covering a 40 arcminute field at the Cassegrain
focus.

Some smaller telescopes offer a larger field of view or larger
multiplex advantage. For example, the 2.5m DuPont telescope at Las
Campanas with a 2.1\degr\ field and up to 200 optical fibres
\cite{lcrs} has been used for the largest previous galaxy
redshift survey. At the 1.2-m UK Schmidt Telescope the FLAIR fibre system
with 92 fibres covered a 6\degr\ field \cite{flair}. FLAIR has now been 
replaced by a semi-automatic successor: 6dF \cite{6df} with 150 fibres. 
The 2.5m Sloan Digital Sky Survey (SDSS) telescope has a 3\degr\ field of view 
and 640 optical fibres; this is a dedicated telescope carrying out
both multi-colour imaging and a spectrosopic survey of a million
galaxies \cite{sdss}.  Multi-object systems are also being built for several 
of the new 8-10m class telescopes; these will be able to observe fainter 
objects but cover much smaller fields of view.

\subsection{Key features of 2dF}

Multi-object fibre systems come in three varieties: plug plates or
other off-line field preparation, for example the SDSS fibre system; 
`fishermen around the pond' type instruments with separate remotely 
controlled mechanisms to simultaneously place each fibre at a target 
position, for example MX \cite{mx} and MEFOS \cite{mefos}; and
`pick and place' systems involving consecutive placement of fibres with 
a single complex robot.
The most versatile are the single robot systems, but they have the
drawback that placing the fibres consumes a significant amount of
observing time.  2dF gets around this drawback by having two independent sets
of 400 fibres and two field plates, mounted back-to-back on a tumbler
assembly within the instrument.  While one set of fibres is being
used for observations, the second set is being reconfigured for the
next target field.

In order to give an unvignetted field of over 2\degr\ with a flat
focal plane, a special corrector lens system had to be designed and
manufactured \cite{commissioning}.

Any wide field multi-fibre system must be able to cope with two
separate atmospheric refraction effects, the variable distortion of the
field as the telescope tracks across the sky, and variation of distortion 
with wavelength, which turns point images into very low
dispersion spectra as zenith distance increases.

\begin{figure}
\centering
\psfig{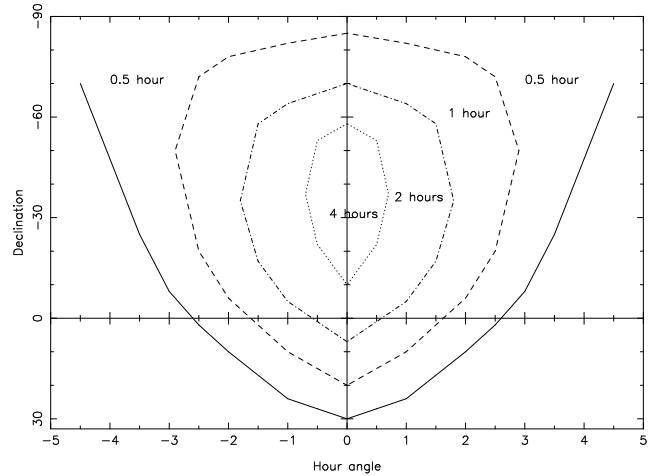}
\caption{ The effect of differential atmospheric refraction at the latitude of the
AAT (-31\degr\ ).
The contours show the maximum possible observation times for 2dF
fields centred at different Hour Angles and declinations, if all
targets are to remain within one third of a diameter of the centre of their fibres.}
\label{darfig}
\end{figure}

The first effect means that for any one target field configuration, there
is a maximum exposure time after which some of
the target objects move out of the fixed fibre apertures.
If we specify that a target field
must be reconfigured once the target objects have moved more than a
third of the fibre diameter (0.7 arcsec in the case of 2dF), this
defines the maximum time available for the robotic positioner to
reconfigure the next set of fibres.  The effect of differential
refraction is dependent on hour angle and declination as shown in
Fig.~\ref{darfig}. Evidently a reconfiguration time of at most an
hour is required for an efficient system which can access most of the
sky with minimal loss of observing time.

The second effect arises because the atmosphere
is a refracting
medium whose refractive index depends on wavelength. For example, at a
zenith distance of 60\degr, the light from a point source with
wavelengths between 365--1100nm is spread out over 4.2 arcsec, which
would make it impossible to use 2 arcsec fibres for low dispersion
spectroscopy over a wide wavelength range. 2dF has an atmospheric
dispersion compensator (ADC) built into the front two elements of the
corrector.  These two elements are slightly prismatic and are
automatically counter-rotated during observations to provide an equal
but opposite dispersion, to counteract the atmosphere as the telescope
tracks across the sky.

The fibres feed a pair of spectrographs which are mounted at the top
of the AAT near the prime focus, to keep the fibres short and maximise
the UV throughput.  Each spectrograph takes 200 spectra
simultaneously, with resolutions of between 500 and 2000, on Tektronix
1024 pixel square CCD detectors.  Mechanisms inside the spectrographs
switch the fibre feeds in phase with the tumbling of the field plates.

All of the hardware making up the 2dF facility, including the
spectrographs and electronics racks, is mounted on a purpose-built top
end ring allowing straightforward interchange with the other three original
alternative top ends of the AAT.  In particular, the entire fibre
system can be left assembled and available for maintenance when 2dF is
not scheduled on the telescope.

\section{Main Components of 2dF}
\label{components}

This section contains technical descriptions of the main components of the 2dF
facility. At the end of this section table \ref{specs} summarises the main instrument properties.

\subsection{Top End Ring}

\begin{figure}
\centering
\psfig{file=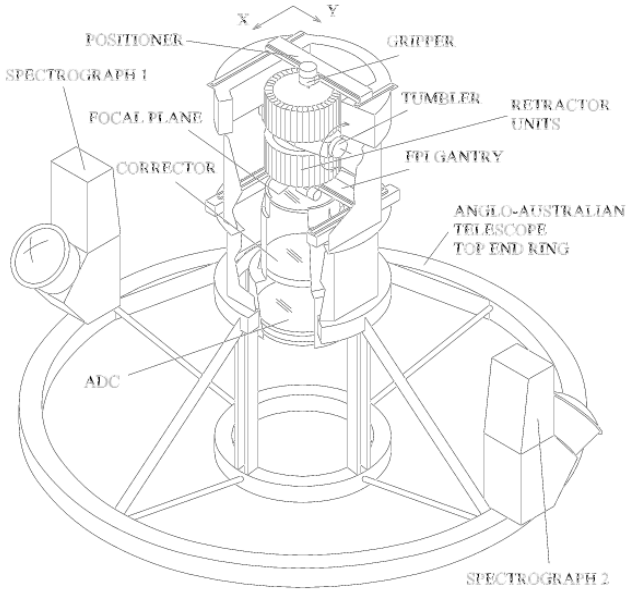,width=\columnwidth,angle=0}
\caption{Schematic diagram of the 2dF top end showing the main components located on the
mounting ring.}
\label{terfig}
\end{figure}

The new 2dF top end ring (Fig. \ref{terfig}) is a direct
copy of the three original AAT top end rings
(f/8 and f/15 + f/35 secondaries and f/3 prime focus) \cite{obsmanual}.
This allows a fast ($<$1 hour) interchange between top ends using the
semi-automated mechanisms built into the AAT dome.

\subsection{Design constraints on the prime focus corrector}
\label{corrector}

   At the heart of the 2dF facility is the corrector lens system which
   provides the 2.1\degr\ diameter field of view at the AAT prime focus. The
   development of a corrector was initiated with a design by
   C.G. Wynne \cite{wynne} offering a 2\degr\ field with 1.5 arcsec
   images using a 4-element corrector. Further work by D. Jones and
   R.G. Bingham emphasised the need for an atmospheric dispersion
   compensator, the importance of chromatic variation in distortion
   (CVD) and of the telecentricity of the optical design. A relatively
   flat focal surface was also a requirement.

   The atmospheric dispersion of uncorrected images when sampled with
   a small fixed aperture size (an optical fibre) will reduce the
   throughput of the system significantly, by an amount which varies
   strongly with wavelength and zenith distance.  When combined with
   small positioning and astrometric errors this will place severe
   limits on the ability to flux calibrate the resulting data. In
   order to minimise this effect an atmospheric dispersion compensator
   built into the corrector optics must provide a variable amount of
   dispersion in the opposite direction to the atmospheric dispersion,
   for as large as possible a range of zenith distances.

   All of the initial designs (except for a significantly aspheric
   design by Bingham) exhibited chromatic variation of distortion
   (CVD) to some extent.  This effect causes off--axis, broadband
   images to be spread radially by up to about 2 arcsec for the 350--1000~nm
   wavelength range, with maximum effect at about 0.5\degr\ field radius.
   This is a smaller effect than that of
   atmospheric dispersion and is independent of zenith distance; it
   determines the ultimate limit to spectro-photometric accuracy with
   2dF.

   The telecentricity of an optical corrector design defines how the
   principal ray of each cone of light reaches the focal plane of the
   telescope. For an ideal fibre system the principal ray should be
   orthogonal to the focal surface. If the input light cone is not
   perpendicular to the focal surface then even in the absence of
   fibre focal ratio degradation (FRD) the effective focal ratio of
   the output beam is decreased. In initial designs the angle of the
   principal ray varied across the field (usually increasing towards
   the edge of the field) by as much as 4\degr\ from the normal to the
   focal plane.  This variation in input angle is effectively the same
   as reducing the input focal ratio of light to the fibre from f/3.5
   to f/2.3. Note that this is a much more severe effect that that of
   FRD within the fibre itself, which is minimal when working at this
   input focal ratio.

   If the spectrograph collimator is oversized to allow for this
   decrease in focal ratio, then we will reduce the spectral
   resolution for a fixed spectrograph beam size and camera focal
   ratio.  Alternatively, if the collimator is sized correctly for the
   f/3.5 beam, fibres accepting light from near the edge of the field
   (the worst affected) will be severely vignetted.  Unfortunately,
   the non-telecentricity of the final optical design means that the
   principal ray varies by up to 4\degr\ from the orthogonal
   case. This effect is partly compensated for in the design of the
   fibre probes (detailed in section \ref{fibres}).  A slight
   oversizing of the spectrograph collimator also reduces the effect
   of the non-telecentricity and any fibre FRD.

\begin{figure}
\centering
\psfig{file=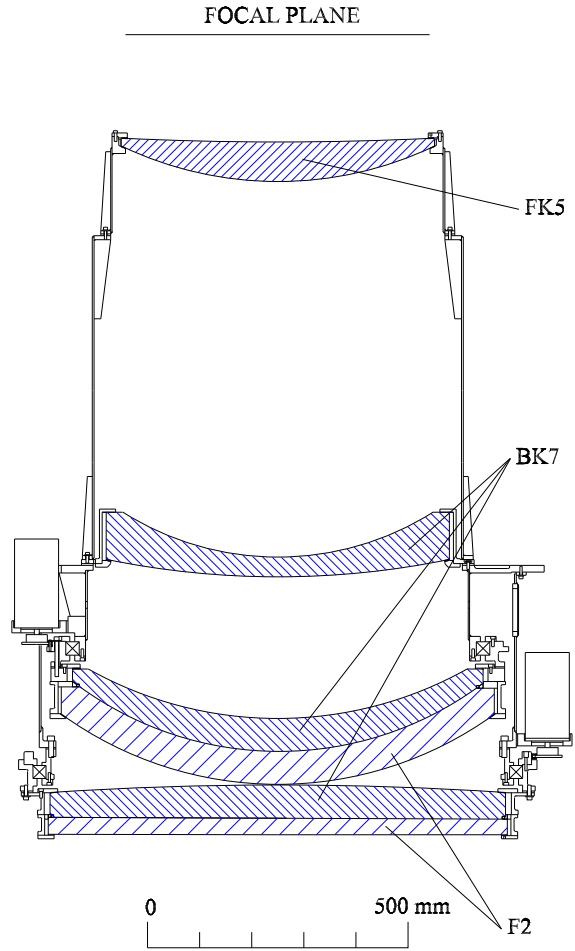,width=\columnwidth,angle=0}
\caption{Schematic diagram of the 2dF prime focus corrector in
cross-section. The lower two lens elements are the prismatic doublets
making up the ADC, these are the first and second elements of the
corrector in the light path.}
\label{corrfig}
\end{figure}

   The design of corrector finally selected \cite{corref} contains
   counter-rotating prismatic doublets as the first two lens elements
   (see Fig. \ref{corrfig}). These provide atmospheric dispersion
   compensation for zenith distances of up to 67\degr. The prismatic
   lenses are designed to give zero deviation of the optical path. CVD
   is maintained below 1.0 arcsec across the field for the maximum
   bandbass used by the spectrographs.  The CVD is zero at the centre
   and edges of the field and reaches its maximum value at 30 arcmin
   field radius.

   The 2dF project is exceptional in that the scientific imperatives
   drove the design towards a large field and high multiplex
   advantage, while the fixed fibre size, realistic astrometric errors
   and relatively poor average site seeing (median value 1.5 arcsec)
   all reduce the importance of the absolute imaging performance of
   the corrector optics. In response to these criteria and the three
   design issues raised above, extensive system modelling was
   performed on the design \cite{modelling}.  This allowed not only the
quantitative assessment of performance issues, but also the evaluation 
of the cost and risk implications of each potential design as work
proceeded.

\subsection{Building the corrector lens}

   At 0.9 metre diameter, the corrector optics contain some of the
   largest refracting elements made for an astronomical telescope.

   The glass blanks for the corrector were manufactured by Ohara
   (Japan). Since some of the corrector lenses are of a deep meniscus
   shape, a technique known as slumping was used to avoid a large and
   expensive wastage of glass and prolonged grinding.  Instead of
   cutting a lens from a thick blank, a thinner glass blank was heated
   and allowed to soften and slump under gravity over a convex
   mould. This technique does have the risk of increased internal
   stress and hence variable refractive index within the slumped
   blank. Tests before and after slumping showed that any variations
   in refractive index were within specification.

\begin{figure}
\centering
\psfig{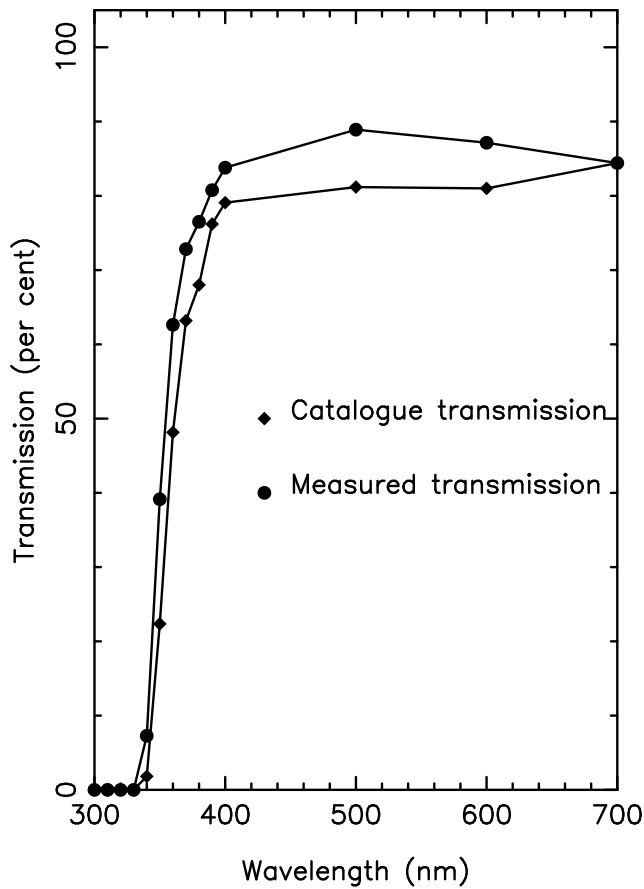}
\caption{The wavelength dependence of the optical corrector transmission.
This is based on measured and catalogue values of the bulk glass transmission
and the predicted transmission of 8 air--glass surfaces with quarter wavelength MgF$_2$
anti--reflection coatings tuned for 500~nm. The actual transmission
exceeds 80 per cent for wavelengths between 385 and 700~nm, falls to 60 per cent at
360~nm and is negligible below 340~nm.}
\label{glasstrans}
\end{figure}

   The internal transmittance, particularly in the UV and blue, was
   also an important consideration in selection of the glass. The
   catalogue specification of UV and blue transmittance of BK7 and F2
   glass types is significantly worse than that of UBK7 and LLF6
   glass. However, discussions with the glass manufacturers revealed
   that the actual melt values were likely to be substantially better
   than the catalogue specification.  Fig. \ref{glasstrans} shows
   the corrector throughput using actual glass melt transmittance measurements,
   compared to the catalogue values for the glasses used in the 2dF corrector.

   The glass blanks were  figured,
   coated with a MgF$_2$ quarter wavelength anti-reflection coating (optimised for
   500~nm) and mounted by Contraves (USA).

   The two ADC elements are prismatic doublet lenses with the BK7 and
   F2 elements in contact to reduce the number of air-glass
   surfaces. An optical coupling compound is used between each of the
   lenses making up the doublets.

   After optical alignment in the corrector cells using temporary
   adjustment screws, the glass elements were mounted using flexible
   silicone rubber which allows for the differential thermal expansion
   of the steel corrector housing and the glass.

   The two lens elements that form the atmospheric dispersion
   compensator are rotated using stepper motors.  These
   automatically move the ADC elements to the required position
   whenever the telescope is slewed and then track continuously during
   observations. A mechanical switch acts as an index mark for each
   ADC element. Step counting is used to measure the position angle of
   the dispersing element. Each long slew of the ADC elements takes up
   to 3 minutes and includes a pass through the index marks to ensure
   that lost steps in the stepper motor do not accumulate and
   contribute to incorrect positioning of the lens elements.

   The optical corrector was received from Contraves as a complete
   unit and commissioned on the AAT between July 1993 and October
   1993, by taking direct night sky images with photographic plates
   and both cooled and uncooled CCD detectors
   \cite{commissioning}. These tests included verification of the
   broad band imaging performance and the operation of the atmospheric
   dispersion compensator, as well as the initial distortion mapping
   of the corrector and telescope optics.

\subsection{Optical fibres and retractor units}
\label{fibres}

\begin{figure}
\centering
\psfig{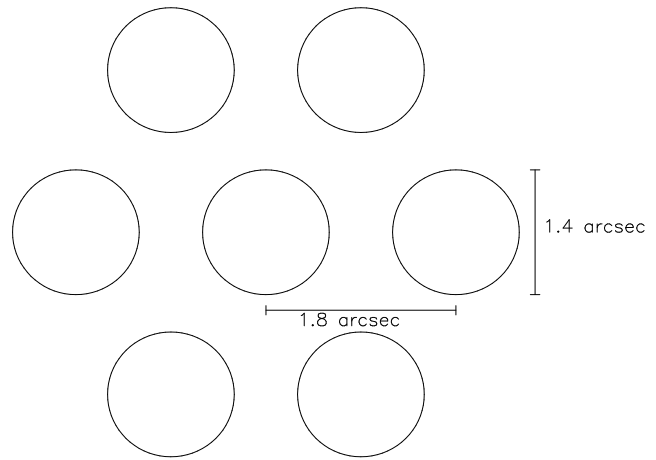}
\caption{Fibre arrangement for a single guide fibre probe showing the
arrangement of the seven individual optical fibres and their size and
separation on the sky}
\label{guidefibres}
\end{figure}

   The focal plane of the telescope is populated by a total of 404
   deployable optical fibre probes which may be moved to cover any
   part of the available field of view of the corrector. The fibre
   probes are divided into two types; one for target objects and the
   other for guide stars. The 400 object fibres each consist of a
   single 8-m long optical fibre of core diameter 140~\micron, 
   corresponding to an average diameter of 2.1 arcsec on the sky (the effective
   diameter decreases non-linearly from 2.16 arcs at the field centre to 2.0 arcsec at the
   edge of the field). The 4 guide fibre bundles each consist of a 4m long 
   coherent bundle of 7$\times$100~\micron\
   core diameter fibres in which 6 fibres are arranged in a hexagon around
   a central fibre (Fig. \ref{guidefibres}). Each individual fibre
   probe can access an area extending from the edge of the focal plane
   to just beyond the centre of the focal plane, and can cover a
   sector with apex angle of 28\degr. For the object fibre probes
   there is sufficient overlap between adjacent fibres to allow full
   field coverage. The four guide fibre bundles can access a total of about
   30 per cent of the focal plane. The guide fibres are arranged at the four
   cardinal points on each fieldplate.

   At the focal plane the incoming light is folded into the optical
   fibres using 92\degr\ prisms (1.2~mm on a side) made from Schott SF5 high
   refractive index glass,
   with the input face anti-reflection coated. The prism is glued to
   the polished end of the optical fibre using UV-curing cement,
   after being optically aligned with the fibre core. The 92\degr\
   angle of the prism was chosen to be half way between the extreme
   ranges of the beam angle for the non-telecentric corrector
   design. The prism material is a high refractive index glass
   so that the fast focal ratio input beam is totally internally
   reflected on the prism hypotenuse. This removes the requirement to
   aluminise the external reflecting face which would result in lower
   efficiency.

\begin{figure}
\centering
\psfig{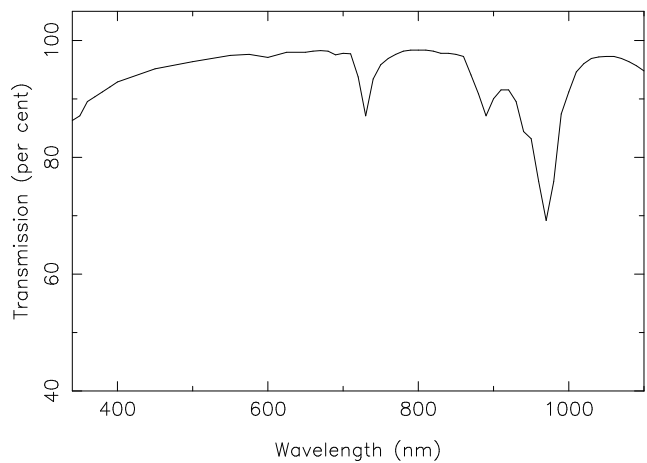}
\caption{Optical fibre throughput as a function of wavelength
(ignoring end losses). Fibre is Polymicro FVP 8m in length.}
\label{fibrethroughputfig}
\end{figure}

   The optical fibre used is a high OH or wet fibre manufactured by
   Polymicro Technologies. This has the advantage of good blue
   throughput at some cost of additional OH absorption bands in the
   far red (see Fig. \ref{fibrethroughputfig}). The optical fibre is a
   step index fibre with core diameter of 140~\micron, cladding
   diameter of 168~\micron\ and a polyimide protective buffer
   198~\micron\ in diameter.

\begin{figure}
\centering
\psfig{file=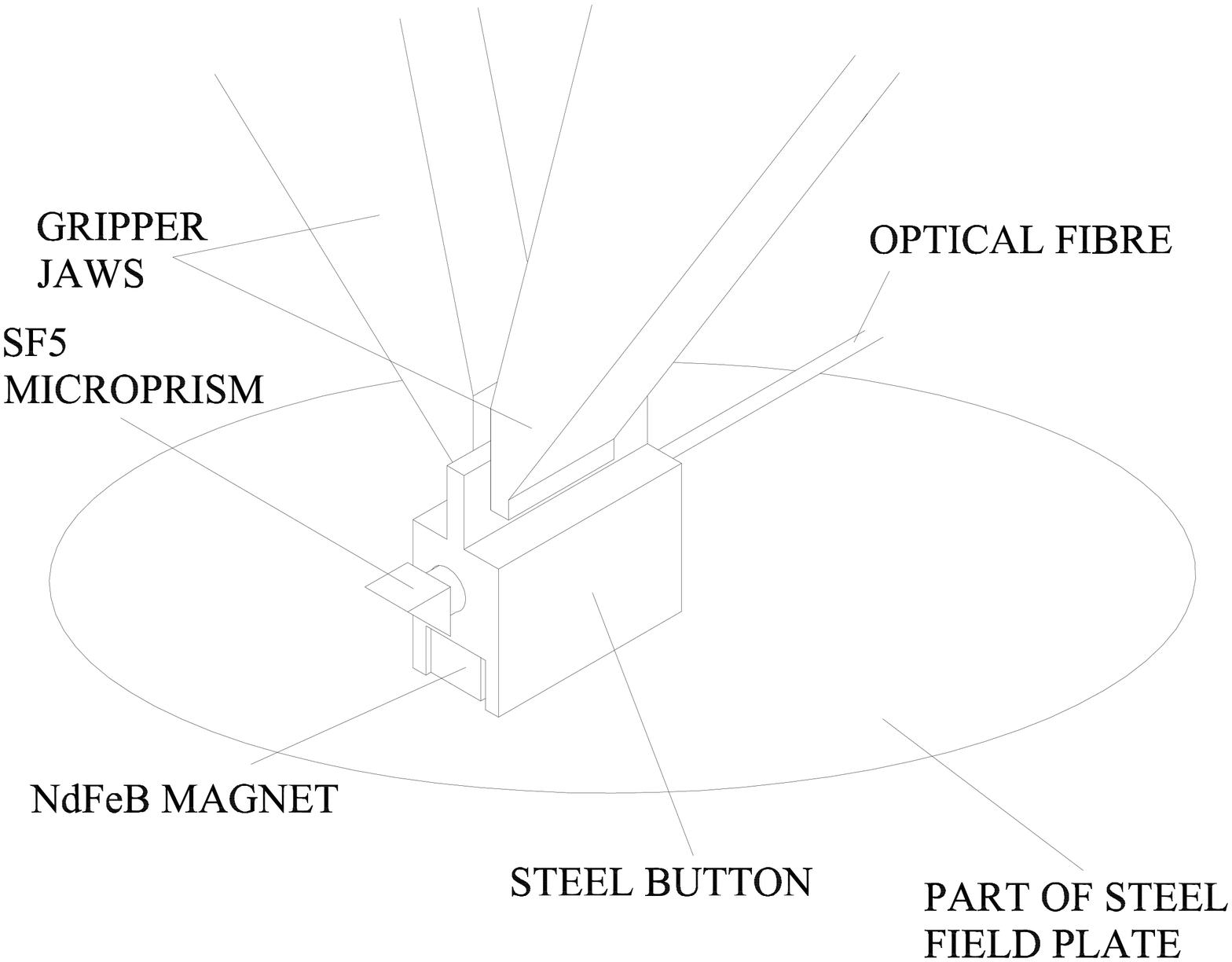,width=\columnwidth,angle=0}
\caption{Diagram showing design of fibre probes and the gripper jaws.}
\label{fibrefig}
\end{figure}

   Each fibre and prism assembly is held in the focal plane by a small
   steel button 4~mm long and 2~mm wide containing a rare earth magnet
   (NdFeB) in its base (see Fig. \ref{fibrefig}).  The magnet holds the
   button in place on a steel field plate located just behind the
   telescope focal plane.  The button has a vertical fin or handle to
   allow the robotic gripper to grasp the button easily. Particular care
   was taken in the design of the button, with extensive simulations of the design
   to minimise the impact of the fibre probe footprint on the success
   rate with which fibres could be assigned to objects
   \cite{simulations}. An overlarge fibre probe footprint would affect
   the success with which fibres could be assigned to target objects
   clustered on relatively small scales, thus potentially imposing an
   instrumental signature on the observations.

   The output ends of the optical fibres at the spectrograph slit are
   aligned together in multiples of 10 fibres which form a
   slitlet. The 10 fibres are aligned and cemented on to a brass block
   using an assembly jig to provide the necessary fibre to fibre
   separation. The fibre separation projects to 5 pixels at the
   detector. The brass block and optical fibres are then polished
   together as a unit.  The slitlets are held in the focal plane of
   the spectrograph as a set of 20 slitlets to form the spectrograph
   `slit' of 200 fibres. The fibres are fanned out to optimise the
   light path through the spectrograph optics.

   Unlike previous AUTOFIB-type fibre positioners, the 2dF fibre probes
   do not have protective stainless steel tubes along their length in
   the focal plane area.  These tubes had a number of uses, they
   protected the fibre against breakages, avoided any risk of fibre
   tangles and simplified the fibre allocation process. However, when
   the fibre probes are
   parked at the periphery of the focal plane the steel tubes must be
   accommodated outside of the fieldplate area which increases the
   overall size of the instrument.  This was not a problem with
   previous Cassegrain focus instruments on the AAT, but would have
   led to loss of light at prime focus.  A second disadvantage with
   steel tubes is that they severely restrict the use of fibre-fibre
   cross-overs in the fibre allocation process, thus considerably
   limiting the number of fibres that can be allocated to target
   objects.

   At the edge of the focal plane the fibres enter
   retractor units which keep the fibres straight on the
   fieldplate. Each fibre is mounted on a pair of pulleys within a
   retractor unit and independently maintained under 30~g tension
   using constant force springs
   \cite{robotvision2}.  After exiting from the retractor unit each
   bundle of 10 optical fibres is protected by a single PTFE tube and
   is routed through the tumbler rotation axis (see Fig. \ref{tumbler})
   and across the
   telescope top end ring spider vanes to the periphery of the top end
   ring where the spectrographs are located. The tumbler rotation axis
   is accommodated by allowing the fibres to twist.

   Each retractor unit contains 10 fibres or 11 fibres (10 target fibres and a
   guide bundle for the 4 retractor units per plate that also carry a guide
   fibre bundle) to
   match the fibre slitlets. This allows for the exchange of a complete
   retractor unit for subsequent repair.  A total of 40 retractor
   units are arranged around the circumference of each of the
   two fieldplates.  This gives a total of 404 fibre probes on each of the fieldplates.
   Ten complete retractor units are maintained as spare
   units.

   This system has the advantage of maintaining the fibres in a
   straight line between the button and edge of the focal plane
   without increasing the footprint of the fibre probe in the focal
   plane.  A drawback is the complexity of
   the retractor units themselves and the high reliability required,
   since a failure of the retractor unit will cause a loop of slack
   fibre to protrude over the fieldplate which may lead to tangles.

   An additional benefit of the bare fibre approach is that it is
   possible for each fibre to cross over many other fibres on its way
   from the edge of the focal plane to its target object. This
   considerably reduces the restrictions on allocating fibre probes to
   objects, at the cost of extra software to ensure that no attempt is
   ever made to move a button while its fibre is overlaid by other
   fibres.

\subsection{Tumbler and field plates}

\begin{figure}
\centering
\psfig{file=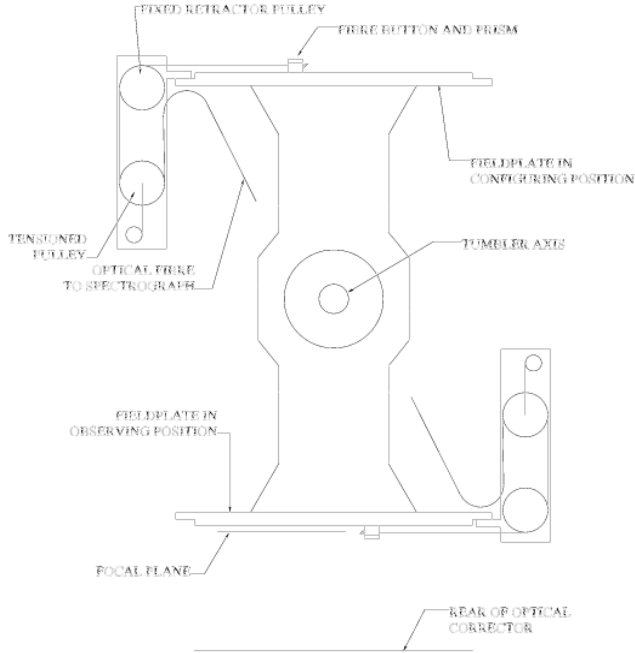,width=\columnwidth,angle=0}
\caption{General arrangement of tumbler unit, fieldplates and fibre
retractors.  Drawing deliberately not to scale to show detail, only
one of forty retractor units shown on each fieldplate for clarity.}
\label{tumbler}
\end{figure}

   The fibres are held in the focal plane of the telescope's prime
   focus by magnetic buttons (as described in the previous section) on a
   flat circular magnetic stainless
   steel field plate 560~mm in diameter. Two of these plates are
   arranged back to back on a tumbler arrangement (Fig.
   \ref{tumbler}). The tumbler can be rotated back and forth through
   180\degr\ to bring either field plate into play at the focal plane
   of the telescope. The other field plate is then ready to be
   accessed by the robotic fibre positioner.

   Embedded in each fieldplate are 21 reference marks in a regular grid
   pattern. These reference marks consist of polished optical fibres inserted into
   holes in the fieldplate. The other ends of the
   optical fibres are illuminated by light emitting diodes (LEDs).
   These reference marks have accurately known
   positions within the fieldplate and are the primary reference frame
   for both the gripper and focal plane imager gantry coordinate
   systems. The use of polished optical fibres means that the gripper optics can
   image and centroid the reference marks reliably.

\subsection{Robotic Positioner}
\label{positioner}

   A single fast robotic positioner \cite{mech} is used to manipulate
   the individual optical fibre probes. Once the robot is given a list
   of X,Y positions (converted from astronomical coordinates provided
   by the observer), the positioner control software works out an
   optimal way to reconfigure the fibres from their current locations
   to the required positions. This usually requires moving a number of
   the fibres out of the way (i.e. parking them at the periphery
   of the field) before moving the majority of the fibres directly to
   their target position. This is at least 50 per cent faster than the more
   simplistic approach of parking all of the fibres before moving each
   fibre to its new target position.

   The robot  consists of a complex gripper unit mounted on an
   X-Y Cartesian gantry,  mounted above the tumbler field plate
   at the top of the 2dF central section (Fig. \ref{terfig}). The
   gripper gantry has two X axes and a Y axis cross-beam in an `H'
   configuration. The axes are driven by linear AC servo motors. The
   use of linear motors over the more conventional leadscrews allows
   faster and more accurate motor control and eliminates the effects
   of backlash. Positional information is determined from independent
   linear optical encoders for each axis, with a resolution of
   1.25~\micron.

   All movements of the XY gantry are balanced by counterweights that
   move in the opposite direction to the gantry, resulting in zero net
   momentum at the prime focus when the positioner is operational.
   With these precautions there is no measureable
   effect on the tracking of the telescope due to the movement of the
   gripper gantry during observations.

   The gripper unit contains a rotational $\theta$ axis to allow the
   gripper to be aligned with the (approximately radially aligned)
   off-axis button handle, a Z axis to raise and lower the gripper
   unit and a set of jaws to grasp the individual button handles. The
   handling of the fibre probes is monitored by a small video CCD
   camera and optics built into the gripper unit. The optical fibres
   are back-illuminated by light projected into the optical fibre
   from the spectrographs. The robotic positioner can `see' the
   illuminated end of the fibre and knows the physical offset to the
   button handle which it grasps. This system allows the gripper to
   monitor fibre placement at the new target position, measure any
   positioning errors and if necessary correct those errors if they
   are deemed to be too large \cite{robotvision1,robotvision2}. This
   has the advantage that the robot positions the fibre core at the
   required position regardless of any actual manufacturing or
   assembly errors in the fibre button assembly. Note that for the guide fibre
   bundles only the central fibre in the coherent bundle of 7 fibres is
   back-illuminated during the positioning process.

   The gripper unit was built under contract by the University of
   Durham to an AAO design, following earlier development work for
   the initial 2dF design study.

   The design of the gripper jaws used in this process is particularly
   important and went through several prototyping stages
   \cite{gripper} in order to achieve repeatability at a level better
   than the overall accuracy requirements. One of the pair of gripper
   jaws is fixed to the XY-$\theta$ gantry and forms a reference
   surface while the second jaw is movable. To grasp a fibre button the
   fixed jaw is positioned against the button handle using the
   XY-$\theta$ gantry, then the movable jaw is closed on to the handle
   to avoid knocking the fibre button unnecessarily. Releasing the
   button requires the movable jaw to be moved away from the button
   handle slightly before the fixed jaw is backed away from the button
   handle using the XY-$\theta$ gantry.

   The robotic positioner is capable of configuring a full target
   field of 404 fibres in about an hour.  This requires on average a
   total of 600 fibre movements to untangle the previous field and
   configure the new target field, at a speed of under 6
   seconds per move.  The average positioning error is 11~\micron\
   (0.16~arcsec ) with all fibres required to be placed within
   20~\micron\ (0.3~arcsec) of the demanded position; if necessary, the
   robot picks up and replaces the button until the position is within
   tolerance.

   The effect of any offset due to flexure between the gripper gantry
   and fieldplate is removed using the grid of reference marks
   embedded in each fieldplate. These are measured with the gripper
   CCD video camera before positioning the fibres on the
   fieldplate. This survey process is always performed after a tumble
   operation or after the telescope has been slewed to a new position,
   before any fibres are moved.

\subsection{Fibre Spectrographs}

   With a fibre-linked spectrograph one has to balance the merits of
   using long fibres to feed a remotely located bench spectrograph,
   against using very short fibres and mounting the spectrographs
   close to the fibre positioner. Each approach has its merits and
   drawbacks. The remotely mounted spectrograph has the advantage of
   providing a stable spectrograph and making the engineering much
   simpler but with the drawback of the lower UV throughput of long
   optical fibres (30 metres). Using short fibres with a locally
   mounted spectrograph reduces the UV light losses due to the fibre
   length but increases the mechanical design difficulties and
   subjects the spectrograph to variable gravitational flexure, and
   ambient temperature variations, as the telescope tracks across the
   sky.

   The choice between a single monolithic spectrograph accepting all
   400 optical fibres compared to a pair of smaller spectrographs each
   accepting 200 optical fibres was largely determined by the
   availability of Tektronix 1024 pixel square CCD detectors.  These
   limit each detector to 200 fibres spaced by 5 pixels on the
   detector. Larger detectors were promised but not available during
   the predicted time frame of the instrument construction.  Limiting
   each spectrograph to 200 fibres also eased the optical design somewhat,
   since it required a shorter slit assembly.

   The decision was made to keep the fibre lengths as short as
   practicable by locating the spectrographs at the top end of the
   telescope \cite{mech}.  Initially they were to be mounted above the
   robot positioner \cite{designsummary} with an extremely short fibre
   length. The final design, however, was constrained by the available
   space envelope and the spectrographs were mounted at the edge of
   the top end ring as shown in Fig. \ref{terfig}. This resulted in
   a fibre length of 8 metres.

\begin{figure}
\centering
\psfig{file=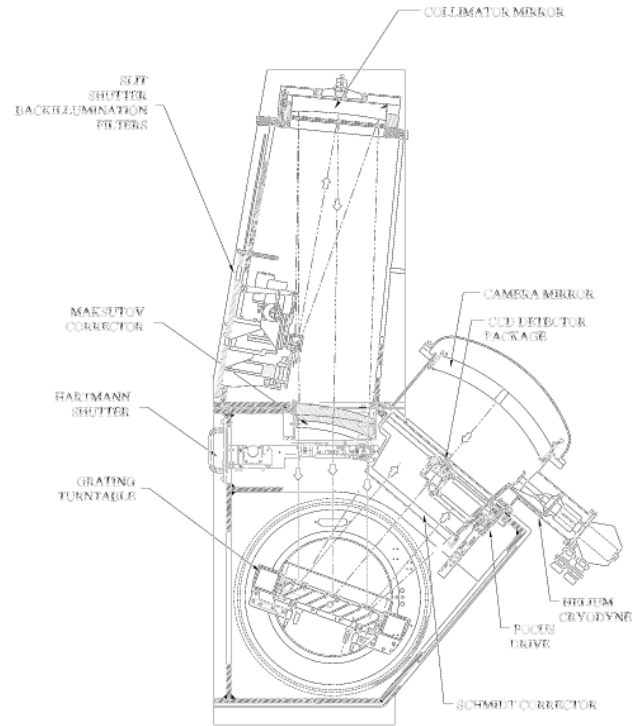,width=\columnwidth,angle=0}
\caption{Spectrograph general arrangement showing the main components
and the optical path}
\label{spectrograph}
\end{figure}

   The overall construction of the spectrographs is depicted in Fig.
   \ref{spectrograph}.  The spectrograph design is based on an f/3.15
   off-axis Maksutov collimator generating a 150~mm beam. This allows
   the use of the existing $150\times200$~mm Cassegrain RGO
   spectrograph reflection gratings. The collimator is slightly
   oversized compared to the prime focus corrector focal ratio of
   f/3.5, to allow for some focal ratio degradation in the optical
   fibres and further reduce any effect of the non-telecentricity of
   the corrector design. The off-axis design has the benefit of no
   central obstruction so the fibre slit and its associated shutter,
   filter wheels and slit interchange mechanism do not vignette the
   beam.

   The 200 optical fibres from the focal plane are arranged in a
   single line to form a pseudo-slit. A shutter is located immediately
   in front of the fibre slit. The shutter is moved in a direction
   perpendicular to the slit axis and therefore only requires a short
   travel to open and close over the slit width as defined by the
   fibre diameter of 140~\micron.

   Two filter wheels, each with four apertures, are provided behind
   the shutter.  One aperture in each filter wheel is kept as a clear
   aperture leaving a maximum of six filter spaces of which three are
   currently used to hold order-sorting filters for high resolution
   observations. The filters themselves are 2~mm thick.
   Filters currently available include
   GG495, S8612 and RG630. Use of a filter requires a change in the
   spectrograph focus as the filters are located in the diverging beam
   behind the fibre slit and shutter.

   The spectrograph camera is located at a collimator-camera angle of
   40\degr\ and is a fast, wide-field Schmidt camera with a CCD at
   its internal focus \cite{camera}. The main difficulty with the
   camera is that the pixel matching requirements of imaging such a
   large number of fibres onto such a small detector force it to have
   an f-ratio as fast as f/1.2 in the spatial direction (f/1.0 in the
   spectral direction) while retaining its wide $\pm 5.5$\degr\ field
   of view. The camera uses a severely aspheric front corrector plate,
   to give good performance over the full wavelength range and field
   angle required to image the spectra.

   The camera is fully evacuated with the aspheric Schmidt corrector
   lens acting as a dewar window. Each CCD is cooled by use of a cold
   finger attached to a CTI cryodyne closed-cycle helium cooler. A
   helium compressor is mounted on the telescope in a gymbal mount and
   is used to drive both cryodynes. The cool-down time for the CCDs is
   relatively long, about 3.5 hours from ambient temperature to their
   operating temperature of 170K. The performance of the cryostats is
   sufficiently high to maintain their performance for up to 21 days.

   In addition to the basic design of the spectrographs, the double
   buffering of the fibre positioner system and the back illumination
   of the fibres provide some extra engineering problems. Each
   spectrograph actually accepts 400 fibres, 200 from each field
   plate, which must be located in the focal plane of the spectrograph
   when required. The non-observing fibres must be illuminated with a
   light source bright enough to be seen by the robotic positioner's
   gripper TV system since fibre positioning is done simultaneously
   with observing. These two requirements mean that each spectrograph
   must have a slit interchange mechanism, a bright light source for
   the fibre back-illumination, and a reliable means of completely
   shielding this light source from the rest of the optical path.

   The spectrograph is focussed by moving the detector within the
   cryogenic camera. Both focus and tilt adjustments of the CCD are
   available and can be driven remotely.

   With the exception of the selection of the grating (which has to be
   manually inserted before observations start), all functions of the
   spectrographs are remotely controlled. This includes slit
   changeover, filters, Hartmann shutter, grating rotation and
   spectrograph focus.  Gratings can be automatically recognised by
   the spectrograph control system which reads magnetic barcodes
   present on all of the grating cells.

   The complete spectrograph configuration is inserted into the data
   FITS header information to allow automatic pipeline data reduction.

\subsection{Focal plane imager}
\label{fpi}

   A second XY gantry, almost identical to the gripper gantry, is
   located immediately behind the final element of the optical
   corrector (Fig. \ref{terfig}) and in front of the observing
   fieldplate. Instead of a gripper unit, this gantry carries a pair
   of CCD cameras. One camera is a simple video CCD camera which can
   only view the back-illuminated fibres and field plate, while the
   second is a Peltier cooled Princeton Instruments CCD camera, facing
   in the opposite direction, which can view the sky when its gantry
   is appropriately positioned. This gantry system is known as the
   Focal Plane Imaging (FPI) system and is the primary means of
   determining the relationship between RA,Dec and X,Y on the field
   plates. This sky viewing camera can also be used for target field
   acquisition and seeing measurements. The image scale of the focal
   plane imager CCD is 0.3 arcsec per pixel.

   The focal plane imager gantry also surveys the reference marks
   embedded in the field plates, in a similar manner to the gripper
   gantry, to correct for registration and flexure before centroiding
   the images of reference stars.

\subsection{Calibration systems}

   2dF contains its own remotely controlled calibration systems. Two
   white flaps may be inserted into the telescope beam below the
   corrector, blocking off sky light and forming a reflective
   screen. A variety of calibration lamps can be used to illuminate
   this screen, with the resulting scattered light travelling through
   the 2dF corrector and illuminating the focal plane. Two intensities
   of quartz lamps for fibre flats and a variety of hollow cathode arc
   lamps (Copper-Argon, Copper-Helium and Iron-Argon) are provided.

   Requesting an arc or fibre flat exposure will result in the control
   system automatically inserting the reflective flaps and turning on
   the requested calibration lamps. At the end of the exposure the
   lamps are turned off and the flaps either removed from the
   telescope beam or left in place if further calibration exposures
   are required. This automation avoids accidentally leaving
   calibration lamps turned on and reduces observing overheads.

   All calibration exposures have appropriate FITS header items added to the
   data frames as an aid to the pipeline reduction of the data.

\subsection{Software Control systems}

   The 2dF facility uses the AAO {\sc drama}  software
   infrastructure \cite{drama} to
   build a fully integrated control system across several computers
   and operating systems.

   The graphical user interface is written using Tcl/Tk and provides
   for control of all aspects of the positioner system, the
   spectrographs, ADC and CCD control, all from one simple to use
   interface.  The same interface allows the observer to control the
   telescope directly, for example when slewing to a new target field.  The
   graphical user interface is provided by several windows spread
   across two computer screens with controls arranged by subsystem.

   Individual parts of the 2dF facility are controlled by software
   tasks which are running on the most appropriate control computer
   (VxWorks, Solaris, VMS) and {\sc drama} allows the many separate tasks to
   communicate and work together seamlessly.

\subsection{CCD operation.}

   The two Tektronix 1024 CCDs are controlled using the standard AAO
   controller hardware, the AAO External Memory (XMEM) and {\sc OBSERVER}
   software interface on a VAX/VMS computer system. With a slight
   adaptation, the {\sc OBSERVER} software communicates with the rest of the
   2dF control system (running on Solaris and VxWorks computer
   systems) using {\sc DRAMA} to enable it to add all of the more
   specialised 2dF FITS header items from the 2dF control system to
   the data files for archiving.  The two CCDs are operated and read
   out simultaneously, but using separate controllers and data links.

\subsection{Acquisition and Guiding systems}

On each fieldplate there are four guide fibre bundles which do not feed to
the spectrographs. These guide fibres consist of a coherent bundle
of seven individual fibres with six fibres in a hexagon pattern
surrounding a central fibre (Fig. \ref{guidefibres}). The
separation of the individual fibres in a guide bundle at the input
end is 1.8 arcsec. Each guide bundle may be positioned on a guide
star in the focal plane to allow for acquisition of the target
field. The output from the four guide fibre bundles is collected by a
Quantex intensified TV system, with its output visible to the
telescope night assistant and observer.  A guide star is accurately
centred when the central fibre of a guide fibre appears brightest
on the TV display and the six surrounding fibres are uniformly
illuminated.

A minimum of two guide fibres are recommended for each target
field, although it is preferable that all four are used in case of
inaccurate stellar positions.  The Quantex TV system can detect
stars down to a magnitude limit of $V=15.5$ in average conditions.
The system was designed to include
an option to rotate the field plates for target acquisition and, if
necessary, during tracking.  However, this has not yet been
implemented.  It turns out that the instrument is sufficiently
stable, and the flexure is small and predictable, so that it is
easier to remove any overall field rotation in the positioner
software.  Often, larger errors would be introduced by attempting
to use the fiducial stars to correct rotation, especially since the
surface density of usable stars means that it is difficult
to obtain enough stars close to the edge of the field to allow rotation to be
determined.

The pointing of the AAT is normally good enough (better than 2 arcsec) that
target field acquisition is a simple procedure of slewing the
telescope and the guide stars will be visible somewhere in at least
one of the guide fibres. Fine acquisition is performed manually by
offsetting the telescope by very small amounts until all four guide
stars are well centred in their respective guide fibres.

If for any reason the guide stars cannot be seen after slewing the
telescope or there is any ambiguity in the guide star acquisition
(for example due to a very crowded field in a globular cluster or
due to poor guide star positions) the focal plane imager CCD camera
can be used to image a small section of sky around each guide star
to verify acquisition. A simple offset will then place the guide
stars onto the guide bundles, after which the focal plane imager can
be removed from the focal plane.

Once a target field is acquired, guiding can be done in one of two
ways.  The normal guiding method is manual guiding using the
excellent tracking of the AAT and making small adjustments every 10
minutes or so.

An automatic autoguider has also been developed which uses the
video output from the Quantex TV system to feed a video frame grabber.
The digitised video can then be analysed to determine the average
centroid of each guide fibre and determine the correct telescope offset 
to restore the telescope pointing. Initially however, the
extra calibration required and problems with the dimensional
stability of the Quantex TV image at different gain levels
meant that autoguiding was not generally used. A simplified 
version of the autoguider system was implemented in 2001 under the control of
the telescope night assistant and this is now in regular use.

\begin{table*}
\begin{minipage}{80mm}
\caption{Summary of 2dF instrument specifications}
\label{specs}
\begin{tabular}{@{}ll}
Telescope                        & 3.9m AAT f/3.3 Prime focus         \\
Corrector field of view          & 2.1\degr           \\
Corrector focal ratio            & f/3.5               \\
ADC recommissioned               & 31 August 1999    \\
Image scale                      & 67\micron/arcsec   \\
Number of spectroscopic fibres   & 400 (on each fieldplate) \\
Number of guide fibres           & 4 (on each fieldplate) \\
Fibre diameter                   & 140\micron \\
Fibre size on sky (mean)         & 2.1 arcsec \\
Reconfiguration time             & 55 minutes (typical) \\
Fieldplate exchange and acquistion & 3 minutes (typical) \\
Robot positioning accuracy       & 11\micron (mean) 20\micron (maximum) \\
Overall positioning accuracy     & 0.3arcsec RMS \\
Length of fibres                 & 8.0m \\
Spectrograph Collimator          & f/3.15 Off-axis Maksutov 150mm beamsize \\
Spectrograph Camera              & Modified Schmidt camera f/1.2 (spatial)  \\
                                 & f/1.0 (spectral) \\
Detector                         & Tektronix $1024\times1024$ pixel CCD \\
CCD inverse gain (NORMAL and SLOW)        & 2.79 -- 1.4 e$^-$/ADU \\
CCD readout times                & 74 -- 120 seconds \\
CCD readout noise                & 5.2 -- 3.6 e$^-$ \\
Pixel size                       & 24\micron \\
Second science grade CCD installed & 31 August 1999 \\
Dispersive elements & Plane ruled reflection gratings $205\times152$mm \\
Range of available dispersions   & 4.8\AA/pixel to 1.1\AA/pixel \\
Range of effective resolutions   & 9\AA~ to 2.2\AA \\
Achievable velocity accuracy${^1}$ & 10 kms$^{-1}$ \\
Spacing of spectra at detector   & 5 pixels \\
Fibre size at detector           & 1.9 -- 2.1 pixels \\
Order sorting filters            & GG495, RG630, S8612 \\
Number of spectra/spectrograph   & 200 \\
System throughput${^2}$          & 5 per cent \\
Minimum object separation in focal plane & 1.6mm (approximately 20arcsec) \\
\end{tabular}

\medskip
$^1$ This is a typical external accuracy for the 1200 lines per mm gratings
and is very dependent on the spectral type of the data.

$^2$ Total system throughput including atmosphere and telescope at 600nm 
with 300B gratings. 
\end{minipage}
\end{table*}

\section{Operation of 2dF}
\label{operations}

\subsection{Input data, astrometry and guide stars}
\label{inputdata}

The successful operation of a robotic fibre system depends
on the provision of accurate positions for the target objects in a
known reference frame.  Target positions are usually measured from
Schmidt telescope photographic plates, using the Automatic Plate
Measuring (APM) or SuperCOSMOS plate measuring machines, although
more recently astrometry from CCD mosaics is becoming more common
and reliable. With care, these provide astronomical coordinates
with a relative accuracy of 0.2--0.3 arcsec across the full
2\degr\ field of 2dF. This involves using a reliable modern astrometric
catalogue  of reference stars and taking full account of proper motions.

To place optical fibres on target objects, 2dF needs to know the
conversion from astronomical coordinates to physical coordinates in
the focal plane. This conversion is determined by measuring the X,Y
positions of sets of astrometric standard stars, using the FPI CCD
camera and sets of fiducial marks on the fieldplate, using the FPI TV camera.
The observed centroids of the star images are matched to
their expected positions, as corrected for atmospheric refraction,
known telescope pointing errors and the distortion introduced by
the 2dF corrector lens.  A least-squares minimisation determines
the fitting parameters; the six free parameters are the scales in X
and Y, the overall rotation and skewness of the field, and any
offset of the centre of the optical distortion pattern from the
optical axis.  Any non-perpendicularity of the axes of the
gantries, or mechanical shifts between the gantries and the field
plates, should be automatically removed by the survey of the
fiducial reference points on the field plates (section \ref{fpi}).

For calibrating 2dF it was not a trivial task to find astrometric data with the
required accuracy (current rms positions to better than 0.25~arcsec
across 2\degr\ of sky) for large enough samples of stars.
Initially, the PPM catalogue was used \cite{ppm}; currently the
Tycho-2 catalogue \cite{tycho2} is the most suitable.  Originally
the astrometric calibration involved taking data for sets of stars
at a range of declinations, for both field plates, and the full
process had to be repeated on the first night of every 2dF
observing run.  Different sets of astrometric parameters were
stored and used as appropriate for each target field.  However, it appears
that the behaviour of 2dF is sufficiently stable and repeatable
that the flexure terms can be predicted.  All that is now needed is
to take at least one set of calibration data whenever 2dF is
re-mounted on the telescope, to determine any rotation or offset
zero-point errors.  There are also small plate scale variations
which are temperature-dependent.  The remaining errors in
the process should be below the 0.5 arcsec level at the edge of the
field, where most effects are worst.

The fiducial or guide star positions
should be as accurate as the target object positions and on the
same astrometric system. However, for a number of reasons this may
not be the case. Firstly fiducial stars are usually in the
magnitude range $V=13-15.5$ so that the Quantex TV system is able to
detect them. Stars as bright as this can suffer halos and
diffraction spikes on the photographic plates which affect their
astrometry. Secondly, proper motions of stars can  increase the
positional errors, particularly when using old plates. Using
fiducial stars towards the faint end of the available range, restricting the colour
range and
comparing two plates of different epochs can reduce both these
effects.  A good solution is to use stars drawn from the target
object list, where feasible. Alternatively, a prescription for the
selection of guide stars is given by Colless \etal \shortcite{grs}.

\subsection{Fibre allocation procedures}

\begin{figure}
\centering
\psfig{file=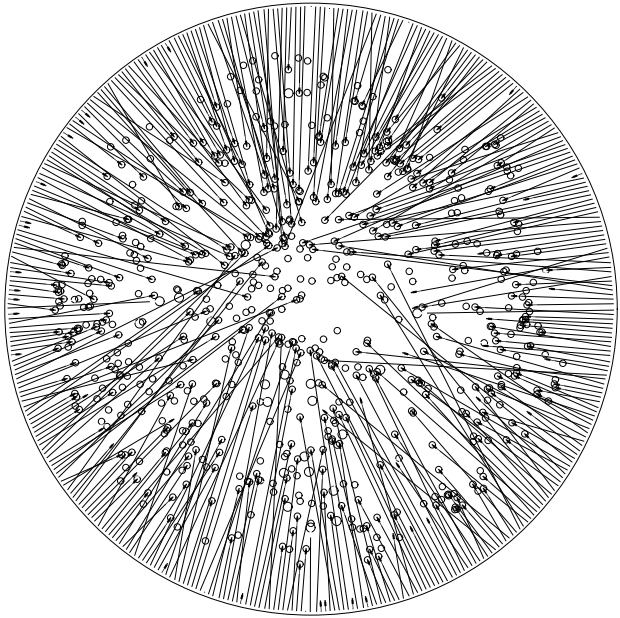,width=\columnwidth,angle=0}
\caption{An example target field configuration using almost all 400
fibre probes.  Unallocated target objects are shown by unfilled
circles.  A few fibres remain parked at the edge of the field because
they cannot access any of the unallocated targets. Small circles represent targets
and large circles are potential guide stars. Fibres apparently on blank areas have
been allocated to sky positions.}
\label{configurefig}
\end{figure}

Fibres are allocated to target objects using an off-line software
program {\sc configure} (see the 2dF WWW pages http://www.aao.gov.au/2df/ for
details of manuals and to download this software)
to preplan each target
field.  An example target field configuration is shown in Fig.
\ref{configurefig}, which is a reproduction of the mimic display at
the telescope.  When provided with a field centre and a list of
target, fiducial star and blank sky positions, the {\sc configure}
program assigns fibres to objects, taking account of the hardware
constraints (limited fibre extension and deviation from the radial
direction) and avoiding illegal hardware collisions.  Given the
size and shape of the fibre buttons, the minimum separation between
two targets is $\simeq 30$ arcsec on the sky, but this is a strong
function of location in the field and of target distribution.

Various options in {\sc configure} allow the user to assign
relative priorities to targets and set other parameters, including
the planned Hour Angle of the observations, in order to optimise
the allocation of fibres.  The program works iteratively making 
a number of random swaps at each iteration to try to
maximise the number of observable targets and minimise the number
of fibre crossovers.  The {\sc configure} utility includes an
option for automatic allocation of fibres to sky positions and
allows manual editing of the final fibre configuration, before
saving as a file to be used with the 2dF control system at a later
stage.  One (or more, if the same field is to be tracked for
several hours) of these configuration files is required for each
target field to be observed during the course of a night.

If the user wishes to observe a field with more than 400
targets, the input list must be split into a series of manageable
field configurations. It is usually best to restrict the magnitude range of
targets in a given field, since bright targets may become saturated
or scatter light on to fainter targets in adjacent fibres in the
spectrographs.

To efficiently cover an area larger than the 2\degr\ field of view
with high completeness requires simultaneous configuration  of multiple fields and adjustment of their field centres, this process is known as `tiling'. A description is the tiling procedure used in the 2dF galaxy redshift survey
is given by Colless \etal \shortcite{grs}.

The offline planning process is usually done well in advance and a
set of configuration files sent to the telescope. Before a set of
configuration files are used they are normally checked for legality
using the current astrometric parameters.

\subsection{Observing procedures}

This section describes the standard observing procedure as used for
the major redshift surveys, although many of the
steps are similar for all 2dF programmes.  Observing with 2dF is
somewhat different from most other observing on the AAT.  The
corollary to the complexity of the instrument, and the need for a
great deal of preparatory work before coming to the telescope, is
that is is difficult to make significant changes to the program
during the night.  Thus there is usually little scope for
interactive decision-making by the observers.

Another fundamental constraint is the fact that it is necessary to
keep planning ahead, since the configuring of a set of targets
takes about one hour, about the same as the time needed for one set
of observations for the redshift surveys.  Thus the observing and
re-configuring have to be carried out simultaneously and almost
continuously, so that any time lost in one aspect will lead to
problems in the other.  Not only that, but the fibres have to be
configured for the expected mean HA of the observations to minimise
atmospheric refraction losses.  One consequence is that if a set of
observations is delayed for any reason, it is rarely possible to
catch up; the best strategy is to abandon that set and move on to
the next set.  Having two field plates also introduces extra
constraints: in the case of the redshift surveys, the
configurations for each field are prepared for one or other
specific field plate.  Thus if one field is lost, it is not
possible to simply move on to the next one; quite often the
consequence of losing one field is an overall loss of two hours of
observing, since the target configurations that have been prepared
are no longer the appropriate ones.

The solution to these constraints is to prepare a detailed
observing plan for each night.  This plan must include all
necessary exposures, including calibrations, with realistic
estimates of the necessary elapsed time for each operation.

Before observing starts, and usually well before sunset, the first
two configurations of the night have their fibres set up on the two
field plates. Depending on the
declination and hour angle, and on the distribution of targets
(centrally concentrated fields can be observed over a much wider
range of hour angle), a target field may have to be observed with multiple
configurations to remove atmospheric refraction effects, as
demonstrated in Fig. \ref{darfig}.

For each target field a minimum set of CCD exposures consists of a
fibre flat (quartz lamp) exposure, an arc exposure and a set of
target object exposures. The fibre flat is used for locating the
spectra on the CCD during the data reduction stage, the arc
exposure is for wavelength calibration of the spectra. Optional CCD
exposures may include offset sky exposures and twilight sky flat
exposures, both used for calibrating the fibre-to-fibre
throughput variation.

If these special exposures are not taken, the fibre to fibre
throughput calibration can often be done using the night sky
emission lines in the object data frames; in fact, there is
evidence that this method actually gives the best results, since
the calibration data are simultaneous with the observations, use
exactly the same optical path and do not involve moving the
telescope.  The night sky line method works best in the
near-infrared part of the spectrum which has many emission
features, or when particularly strong night sky lines are present,
such as [OI] at 5577\AA\ and 6300\AA\ \cite{2dfdr2}.

Between target fields, the fieldplates must be exchanged using the
tumbler, and the telescope and dome must be slewed to the new
target field position.  The ADC automatically tracks the telescope
position thus removing this complication for the observer. A
straight-forward field changeover takes under three minutes. Once
the telescope is tracking at the new target field location the
robot positioner can start moving the fibres for the subsequent
target field.

Observations of single standard stars (radial velocity standards,
metallicity standards etc) are possible without reconfiguring all
of the fibres. A calculation tool is provided to allow blind
offsetting of a standard star into a single spectroscopic fibre,
using a nearby guide fibre.  This process is however not suitable
for observing flux standards, due to the uncertain nature of the
final blind offset and the small size of the fibre aperture.

It is not yet clear how accurately it is possible to calibrate the
fluxes in 2dF spectra, especially when taken over a wide wavelength
range.  Absolute fluxes can never be very accurate, since the fibre
diameters (2 arcsec) are comparable to the typical `seeing' disk
size, and the overall accuracy of fibre positioning is often no
better than 0.5 arcsec.  Relative fluxes, i.e. as a function of
wavelength, can be determined to higher precision.  However, here
the CVD inherent in the 2dF corrector lens (see section \ref{corrector}) limits
the accuracy attainable.  Repeatable results should be obtainable
for stars near the field centre, where the CVD is small, but over
most of the field the combination of fibre positioning errors,
CVD and variable seeing means that the slope of stellar spectra varies 
substantially, even between consecutive short exposures \cite{calib}.

\subsection{Data reduction software}

During the design stages of 2dF it was clear that a dedicated
pipeline data reduction package would be required to deal with the
enormous amounts of data, up to 3000 spectra, which 2dF would be
capable of generating during a single night.

A completely new data reduction software package ({\sc 2dfdr}) has
been developed for 2dF \cite{2dfdr1,2dfdr2}. The design philosophy
of the {\sc 2dfdr} package is to make as much use of known
instrumental parameters as possible.  In particular, {\sc 2dfdr}
uses the known optical properties of the spectrographs to predict
the format of the data (location of the spectra and dispersion).
The {\sc fits} headers of the data contain all the information
needed to identify the different types of data frame, the location
of all the target objects, details of the telescope and
spectrograph parameters, and the appropriate calibration frames to
be associated with each data frame.  The {\sc 2dfdr} package
preserves the integrity of the data headers, together with the
statistical variance of each spectrum and the sky spectrum that was
associated with each object.

The {\sc 2dfdr} program can be run automatically and is able to
reduce an entire night's data in an hour or less.  However, it is
normally better to do at least some of the steps interactively,
especially the initial determination of the `tramline' extraction
maps for the spectra, and to verify that the results are sensible
at each stage.  The software package provides several options and
many parameters which can be set, and includes a number of
diagnostics for assessing data quality.

The data for a particular target field are reduced as two separate
sets of data, one for each spectrograph containing 200 individual
spectra.  For the large redshift surveys, it is normal procedure to
analyse the data as the night proceeds, with all observations fully
reduced and all redshifts determined by the end of the night's
observing.

The data reduction software package {\sc 2dfdr} is available for
download from the 2dF WWW pages (http://www.aao.gov.au/2df/).

\section{Performance of 2dF}
\label{performance}

\subsection{Corrector optics}
\label{opticsperf}

The initial imaging tests \cite{commissioning} showed that the 2dF
corrector lens optics met their design specification.  The imaging
performance was also checked in May 2001, after some evidence of apparently
variable throughput across the 2dF field (see section \ref{calibperf} below).  A
series of spectra of a set of standard stars was taken, stepping through
focus between exposures.  These were somewhat inconclusive, given the
fibre diameter of 2 arcsec and the sensitivity to `seeing' variations.  A
better check was provided by taking a series of exposures on a sheet of
photographic film, stuck on a sheet of magnetic material which was then mounted on one
of the field plates.  These showed that the focus is constant across the
2dF field to within 30 microns; there is no tilt of the focal surface and
no serious variations in PSF across the full field.

\subsection{Atmospheric Dispersion Compensator}
\label{adcperf}

\begin{figure}
\centering
\psfig{file=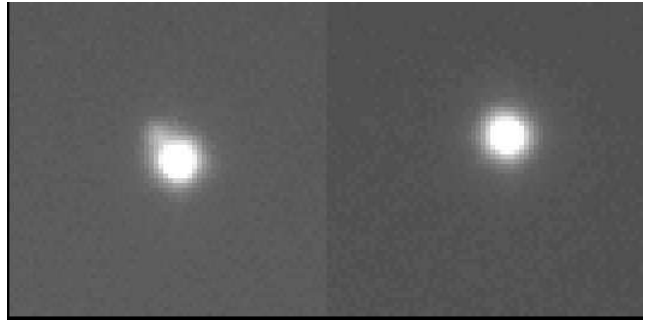,width=\columnwidth,angle=-90}
\caption{Operation of the ADC at 67\degr\ zenith distance. The left hand image
shows the image quality with the ADC nulled, the right hand image shows the same
stellar image with the ADC tracking at the appropriate orientation. Both images
were taken in 1.5 arcseond seeing. The left hand image shows the faint UV
stellar image offset to the upper left of the brighter red image. Each image is
66 pixels (20~arcsec) on a side.}
\label{adctest}
\end{figure}

The effectiveness of the ADC has been
demonstrated using the focal plane imager CCD camera to take an image of
a star at a zenith distance of 67\degr. A filter glass (Schott BG1) was 
inserted in the optical path which absorbs the visible light and passes UV and near--IR. The results
are shown in Fig.
\ref{adctest}. The image taken with the ADC nulled (no correction of
atmospheric dispersion) shows a slight separation of the UV and near IR
images, the image taken with the ADC operating shows a circular
superimposed set of images.

The ADC was was shown to work well for imaging during the corrector lens
acceptance tests in 1993. For these tests the ADC elements were manually 
driven to the correct position. However, later checks on the spectra of some
bright stars at large zenith distance revealed a decrease
instead of an increase in UV flux when the ADC was activated.  This turned
out to be due to a combination of software and zero-point errors in
driving the two counter-rotating elements of the ADC.  Since the pointing
and guiding of 2dF is done at an effective wavelength of about 500nm, this
means that all low-dispersion 2dF spectra taken up until August 1999 are
liable to have been degraded by an effectively random loss of UV and (to a
much lesser extent) near-IR flux, even for fields observed near the
zenith.  The effect is less serious for most high dispersion spectra since
they cover a shorter wavelength range. This problem was corrected in 
August 1999 and all subsequent data have the ADC operating correctly.

\begin{figure}
\centering
\psfig{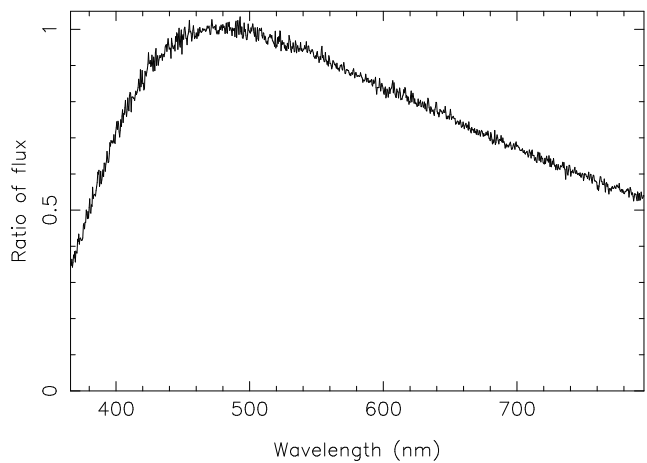}
\caption{The ratio of two spectra of a star observed at a zenith
distance of 50\degr, one taken with the ADC tracking and the other
with it nulled.  In this example, about 65 per cent of the UV light
and 50 per cent of the near infrared flux missed the 2 arcsec diameter
fibre when the ADC was nulled.  See text for details.}
\label{adcspectra}
\end{figure}

Direct tests on stellar spectra confirm the effectiveness of the ADC,
which can increase the UV throughput by factors of 3 or more \cite{calib}. 
Fig. \ref{adcspectra} shows the ratio of two consecutive short exposure
(5s) spectra of a bright hot star, observed in good seeing (0.9
arcsec) at a zenith distance of 50\degr; a spectrum taken with the ADC
nulled has been divided by an otherwise identical spectrum taken with
the ADC tracking.  The star was the ninth magnitude Wolf Rayet star
HD76536, which has broad emission lines up to ten times stronger than
the continuum.  The virtually perfect cancellation of these dominant
features demonstrates that the spectra have been well calibrated in 
wavelength.  In
this case, the spectrum taken without the ADC misses up to 65 per cent
of the UV flux and 50 per cent of the IR flux.  The numerical losses
depend strongly on the effective wavelength at which the uncorrected
image is acquired, the seeing and any positioning errors, as well as
the zenith distance.

\subsection{Positioner}
\label{positionerperf}

The accuracy, speed and reliability of the positioner are all crucial for
efficient operation of 2dF.

The positioner internal precision is set to 20 microns (0.3 arcsec) by
requiring that the robot keep repeating the placement of each button until
the actual position is within 20~\micron\ of the demanded position (section
\ref{positioner}); normally the average number of iterations per fibre move is much
less than unity.  The external accuracy actually achieved is much harder
to assess since it depends on many factors: the reliability of the input
target positions (which in turn can involve several effects such as proper
motions, magnitude or colour-dependent astrometric errors, and inaccurate
transformations from some original x,y coordinate system to RA and Dec);
the accuracy with which the 2dF gantry system has been calibrated
(section \ref{inputdata}); and the stability of the robot during the hour or so
required to do a full configuration, which itself may be a function of the
zenith angle of the telescope (since configuring is usually done while
observing the previous field).

The overall positioning accuracy of 2dF can be assessed in three ways: by
how well the four guide stars can be acquired simultaneously, by comparing
the counts in each fibre with the known magnitudes of the target objects,
or by taking offset exposures to determine the actual positions of targets
relative to the fibres.  The first two ways are used routinely during
normal observations and can quickly show if there is a problem; the last
is a time-consuming procedure used only for tests of the system.  If the
guide star positions appear to be inconsistent, a simple check can be done
by asking the FPI imaging CCD to determine the centroid for each star.

\begin{figure}
\centering
\psfig{file=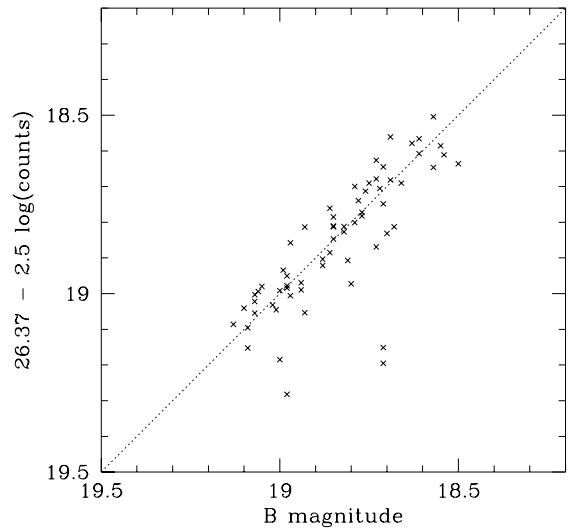,width=\columnwidth,angle=0}
\caption{Counts versus magnitude plot for a sample of 67 main sequence stars in
the globular cluster Omega Centauri, from a total exposure of 10 hours
per star in the blue spectral range.  The zero point of the ordinate
has been chosen to give the best fit to a 45\degr\ line after
discarding 3 discrepant points.  The remaining stars show an rms
scatter of 0.07 mag or about 7 per cent in flux; the component of this
scatter due to positioning errors is at most 5 per cent.
}
\label{counts_mag}
\end{figure}

The best tests of the system are observations of moderately bright ($V
\sim 15$) stars in Galactic clusters or the Magellanic Clouds; such
samples of stars are often restricted to narrow ranges of magnitude and
colour, and all have virtually identical proper motions.  Sometimes the
guide stars can be selected from the same target sample.  In the best
cases, all four guide stars will be well centered simultaneously and the
plots of counts versus magnitude (available routinely within the {\sc
2dfdr} reduction software) will show a scatter of no more than
$\pm0.2$~mag, implying that all fibres have been correctly placed to better
than 0.5 arcsec. An example of this is shown in Fig. \ref{counts_mag}.  The fact that such good results are obtainable indicates
that the inherent accuracy of the 2dF robot is at that level, and that
most of the worse results must be a consequence of either bad input data
or a poorly calibrated positioner.  In general, the counts versus
magnitude plots have larger scatter, particularly for targets near the
edge of the field, and there are often a few outliers with very low
counts, corresponding to position errors larger than the 1 arcsec radius
of the fibres. Of course the counts versus magnitude plots show much greater
scatter for extended galaxies, since a variable fraction of the total galaxy
light enters the fibre, and the astrometric tolerances for such targets can be
somewhat looser. Work is continuing on the precise calibration of the
positioner, and on determining the best values of parameters such as the
temperature dependence of the plate scales.

The speed of the positioner is approximately 6 seconds per fibre move, so that a
full re-configuration of all 400 fibres (requiring $\sim600$ moves) takes
about one hour.  The original 2dF specification was to take only 30-40 min
per reconfiguration, but since an elapsed time of close to an hour was
found to be optimal for the redshift surveys (allowing for the
necessary calibration exposures and CCD read-outs) this was relaxed
somewhat.  The original specification was driven by the need to
re-configure fields to counter the effects of atmospheric refraction, but
this has turned out rarely to be an operational constraint since the limiting
time would only be an hour at rather extreme zenith distances.  Observing
at large zenith distances is discouraged in any case, since the positioner
accuracy and reliability deterioriate.

The reliability of the positioner was originally specified to be no worse
than $1:10^{5}$ per fibre move, on the grounds that a failure was liable
to be catastrophic and could lead to the loss of hours of observing time
if there was a major fibre tangle or many breakages.  Such serious
failures must not occur more than once per lunation.  Initially, when 2dF
observations began, the failure rate was much worse than this.
However, the incidence of catastrophic failures has become very low as
various improvements have been made to the hardware and control software.
The current rate of positioner failures of all types averages approximately 
$1:10^{4}$ fibre movements, but most failures are relatively trivial, cause no
serious damage and cost only minutes to fix.

\subsection{Optical fibres}
\label{fibreperf}

The variations in fibre throughput are small.  The overall scatter, as
measured from the throughput calibrations determined by {\sc 2dfdr}, is less
than 20 per cent for most of the fibres and this includes the variations of
15 per cent in fibre cross-section from field centre to edge, plus any variations
in the prisms at the input ends.  A few fibres have lower transmission by
factors of about two, presumably due to misaligned prisms, dirt or other
accidents (parked fibres appear to have systematically lower transmission,
less than half the normal value, but this is simply due to vignetting).
The throughputs appear to be stable with time.  There is little evidence
for differences in fibre transmission with wavelength: any such variations
are essentially trivial, compared with the differences in spectral shape
introduced by the radial variations (CVD) intrinsic to the 2dF optics
(see section \ref{corrector}) combined with fibre positioning errors.

Occasionally, spectra appear with a marked saw-tooth intensity variation
of amplitude up to 20 per cent;  the number of oscillations can range
from two or
three to a dozen or more, with the period of the variations
increasing linearly with actual wavelength.  These appear to be an effect
of interference fringing, probably due to cracks in the fibres or some
fault at their terminations.  The variations are usually removed or much reduced by
the standard flat-fielding procedure in {\sc 2dfdr} but this sometimes fails,
either because the fibres are unstable or through some error in the flat
fielding process.

\subsection{Spectrographs}
\label{spectrographperf}
   The optical quality of the spectrographs is such that each fibre
   image is focussed to a resolution of about two pixels.  Typical
   values of arc line widths are a FWHM of 2.1 pixels for spectrograph
   No. 1 and 1.9 pixels for spectrograph No. 2.  These values yield
   resolutions with the various gratings as shown in Table
   \ref{gratings}.  There is some variation in spectral resolution
   across the CCDs, due to a deterioration of image quality in the
   corners of the field; this affects in particular the spectral
   resolution at the UV and near-IR ends of spectra in the first and
   last slit blocks (spectrum numbers 1-10 and 191-200).  This
   variation in PSF across the field sets one limit to the accuracy of
   sky subtraction; even if the intensities are calculated correctly,
   the changes in line profile lead to residual P Cygni-type dips and
   spikes at the locations of strong night sky emission lines.  The
   change in image quality also leads to poorer spatial resolution, and
   hence to occasional problems with fitting the tramline map which is
   used to extract the 1-D spectra.

   The optical imaging quality of the spectrographs does limit their
   usefulness in some ways.  There is significant astigmatism at high
   grating angles, making it impossible to use them effectively in
   second order and degrading their performance with the 1200 lines per mm
   gratings at the near-IR end of the wavelength range (e.g. for the
   Ca II triplet near 8600\AA~).  It is possible to set good focus in
   either the spatial or the spectral direction, but not both
   simultaneously.

   The focus of the spectrographs is quite strongly dependent on
   temperature, sometimes necessitating re-focussing during the course
   of the night if the ambient temperature changes by more than a few
   degrees.

   Flexure in the spectrographs is at the level of $\sim0.2$ pixels
   per hour. For most observing programs, individual
exposures are for at most 30 or 40 minutes in order to control cosmic ray
contamination and it is standard practice to take calibration arc
exposures every hour or so.  For the main
   redshift surveys, which are done at low resolution with the 300B
   gratings, the accuracy of velocity determinations for galaxies
   (determined from repeat observations of the same targets) is about
   65 kms$^{-1}$ for the best quality data and around 100 kms$^{-1}$ on average
   (Colless et al, 2001).  Much higher accuracies are attainable for
   stars observed at high resolution with the 1200 lines per mm gratings:
   typical values are in the range 5-10 kms$^{-1}$.

   The throughput of the spectrographs is reasonably uniform across
   the full field of view, with a scatter of about 10 per cent with no drop off
   towards the edges of the CCD, much superior
   to the earlier AUTOFIB system when used with the RGO Cassegrain
   spectrograph.  Scattered light is however a significant problem
   which manifests itself in various ways.  Direct measurement of
   scattered light can be done in the spectral direction by looking at
   the profiles of isolated strong arc lines, or in the spatial
   direction by illuminating a single fibre.  Alternatively, the total
   scattered light in a data frame can be estimated from the residual
   signal in broken or masked-off fibres.  All these methods give
   consistent results.  The total amount of light scattered from a
   single illuminated pixel to other pixels on the CCD is between 10
   and 20 per cent of the input signal; the exact amount varies with
   wavelength and location in the field.  Another measure is that
   about 1 per cent of the light in a single fibre is scattered into each of
   the immediately adjacent fibre locations on the CCD (note that this
   virtually all occurs within the spectrograph,  there is no detectable
   `cross-talk' between fibres before they reach the spectrographs).
   A practical consequence of this is that it is inadvisable to try to
   observe stars with too wide a range in magnitude simultaneously,
   and it is important to avoid accidentally hitting any bright stars
   when observing faint targets.

   Correction for any scattered light is important for the proper
   determination of relative fibre throughputs and for accurate sky
   subtraction.  It is also of course essential for the measurement of
   the equivalent widths of absorption features.  For fields of
   relatively faint targets, or whenever the range of signal strengths
   in different fibres is small, the total scattered light
   distribution can be approximated by a smoothly varying function
   across the CCD.  This is the approach used in the redshift surveys
   and is the default option within {\sc 2dfdr}; a mean background map
   is determined by fitting to the signal in broken fibres and at the
   edge of the frame, and this is then subtracted from the data
   \cite{2dfdr2}.  The accuracy of sky subtraction in the redshift
   surveys is typically 2--5 per cent, although much worse
   values are sometimes reported.

   A special type of scattered light arises in the case of halation,
   which is sometimes a serious problem especially in spectrograph camera
   No. 2.  A condensate slowly builds up on the field flattener lens
   in front of the CCD and causes the formation of haloes around
   bright sources.  These usually appear as shoulders or a pedestal at
   the base of strong emission lines, or adjacent to bright stars in
   the spatial direction.  In some early 2dF data this halation could
   contribute a further 10--15 per cent to the total scattered light.  This
   then substantially degraded the data, since it gave rise to both a
   loss of signal (the total amount lost to scattering must be
   considerably larger than the numbers reported here, which refer
   only to the light which lands on the CCD) and an increase in noise.
   However, it can be kept to much lower limits by careful monitoring
   and regular pumping of the cryogenic CCD cameras.

   The level of light leaking from the fibre back-illumination within
   the spectrograph is a potential cause for concern. This can be
   tested by illuminating the fibres continuously while taking a dark
   CCD exposure and measuring the excess signal over a similar length
   exposure with no fibre back illumination. This has to be repeated
   for both spectrographs and both plate combinations. Typically $<$1
   ADU of additional diffuse background signal is measured from the
   permanently back illuminated case in a 5 minute dark exposure.  In
   normal use the back illumination is only switched on when the
   positioner robot requires to see the fibre. This results in a duty
   cycle of less than 10 per cent so the light leaking from the back
   illumination is insignificant for the typical length exposures of
   20--40 minutes.  True dark current in the 2dF CCDs is at a
   negligible level for most purposes.

   The quality and stability of the CCDs and controllers are very good 
   and there is
   normally no need to take bias frames. The bias counts are well subtracted by
   using data from the overscan region of the data frames.

   Details of the characteristics and performance of the CCDs can be
   obtained from the 2dF WWW pages (http://www.aao.gov.au/2df/).  
   The two Tek 1024 CCDs in use in
   2dF are very similar in all respects.  Prior to September 1999 an
   engineering-grade CCD was in use in spectrograph camera No. 2; this too had
   similar characteristics, but with a few bad columns and some other
   cosmetic defects.

\begin{table}
\caption{Properties of pairs of gratings available for use with 2dF}
\label{gratings}
\begin{tabular}{@{}lcccccc}
Grating  & $\rho$ & $\lambda_b$ & dispersion (nm/pixel) & resolution (nm)  \\
300B     & 300    &  420        & 0.43                  &    0.90	   \\
270R$^1$ & 270    &  760        & 0.48                  &    1.0           \\
316R$^1$ & 316    &  750        & 0.41                  &    0.85          \\
600V     &  600   &  500        & 0.22                  &    0.44          \\
1200B    & 1200   &  430        & 0.11                  &    0.22          \\
1200V    & 1200   &  500        & 0.11                  &    0.22          \\
1200R    & 1200   &  750        & 0.11                  &    0.22          \\
\end{tabular}

\medskip
$\lambda_b$ is the Littrow blaze wavelength. The resolution is the effective
instrumental resolution, either the projection of the fibre onto the detector or
2 pixels, whichever is the larger. $\rho$ is the ruling density (lines per mm).

$^1$ The 270R grating is paired with a 316R grating since the 270R grating
master was no longer available when a duplicate was ordered.
\end{table}

\subsection{Data reduction and calibration}
\label{calibperf}

The pipeline data reduction system {\sc 2dfdr} is
a crucial component of 2dF, enabling observers to assess data quickly
during the night and to come away at the end of an observing run with
fully reduced data.  This was essential for the redshift surveys, where
it was necessary to keep pace with the incoming data and the
quasi-on-line data reductions had to be the final output.  For most other
projects, the many options within {\sc 2dfdr}  make it a very powerful and
convenient tool for producing high quality reduced data, providing it
is used carefully and with a full understanding of the procedures. 

Aspects of the performance of {\sc 2dfdr} are discussed in the 2dF
User Guide (http://www.aao.gov.au/2df/), by Bailey \etal \shortcite{2dfdr2},
by Colless \etal \shortcite{grs} in the
context of the redshift survey and in preceding sections of this
paper.

There remain some situations where it is not initially clear whether a
particular problem is a feature of the 2dF instrument itself or an
artefact of the data reduction process.  A good example was afforded
by the demonstration by Croom \etal \shortcite{qrs} that there are apparent
systematic variations in throughput across the 2dF field.  Their
fig. 5 is a contour plot, showing the difference between the actual
counts obtained and the expected magnitudes for over 20,000 quasar
candidates in the 2dF Quasar Redshift Survey survey.  It appears that the
2dF throughput is
systematically low at the western edge of the field, and that the peak
throughput is offset towards the eastern edge, well away from the
field centre; the amplitude of the effect is $\pm 0.4$~mag or about a
factor of two.  This result led to a rapid checking of the astrometric
and imaging performance of 2dF, but it seemed that neither positioning
errors nor variations in focus or image structure could explain the
effect.  A similar effect, but at somewhat lower amplitude, was seen in
the corresponding  galaxy redshift survey data.

It transpired that the cause lay in a partial failure of the
sky subtraction procedure, combined with some details of the way in
which the software calibrates the relative fibre throughputs.  Spectra
near the edges of the CCD frame in the 2dF spectrographs have slightly
poorer resolution than the majority; this led to an underestimate of
the strength of night sky lines in the spectra and an over-correction
of the fibre throughput.  Both effects led to an artificial increase
in the photon counts for spectra near the edges of the CCDs, which
happen to come from fibres which populate the eastern part of the 2dF
field; conversely, spectra near the centre of the CCD which correspond
to the western part of the field (and which are actually the best
quality spectra) appeared to be giving fewer counts.  Correction of
the original sky subtraction error led to a much more symmetrical
plot of counts versus magnitude, in which the peak is at the field
centre and there is some decrease all round the periphery, as expected
given the larger astrometric errors and somewhat worse imaging towards
the edge of the 2dF field.

\subsection{System throughput}

The total system throughput (including atmosphere, telescope, fibres, 
instrumentation
and detector quantum efficiency) has been determined by observing fields of
several dozen photometric standard stars to allow an average to be determined
over a large number of fibres. With the 300B grating we measured the absolute
efficiency (corrected to 1.0 arcsec seeing) with typical values as given 
in Table \ref{efficiency}.

\begin{table}
\caption{Broadband total efficiency measurements}
\label{efficiency}
\begin{tabular}{@{}lcccccc}
Passband      &  wavelength  & electrons/s/\AA & efficiency  \\
and magnitude &     (nm)     &                 &  (per cent) \\
B = 17        & 440          & 0.6	       &    2.8      \\
V = 17        & 550          & 0.6	       &    4.3      \\
R =17         & 700          & 0.4	       &    4.7      \\
\end{tabular}
\end{table}

\begin{figure}
\centering
\psfig{file=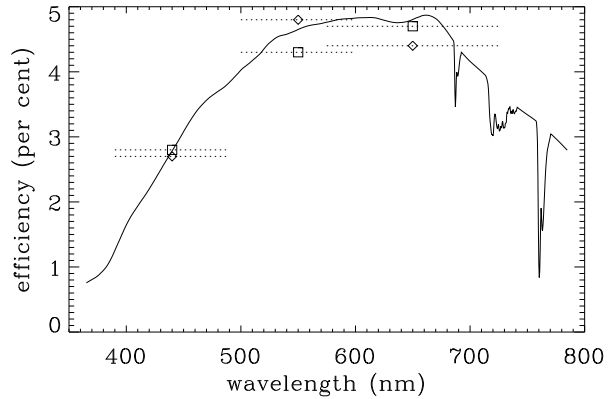,width=\columnwidth,angle=0}
\caption{Figure showing the total 2dF system efficiency with 300B gratings as a function of wavelength. The
squares and diamonds represent broadband absolute efficiency measurements from 1997 January
and 1999 November, respectively. The solid line was obtained from a relative throughput
calibration using data from 2001 January scaled to approximately match the broadband
measurements.}
\label{2dfefficiency}
\end{figure}

Since the wavelength information in these broadband measurements
was insufficient, an improved relative throughput calibration was derived
using about one hundred stars observed simultaneously with 2dF (Fig.
\ref{2dfefficiency}).
The relative efficiency versus wavelength was determined by dividing the
observed spectra by model spectra that were chosen to best match
the stars' broadband magnitudes.

Of course these efficiencies change with choice of grating, the best
performance comes with the use of the low resolution 270R and 316R
gratings where the blaze of the grating matches the peak of the CCD
quantum efficiency. In this case the system efficiency peaks at 9 per
cent. For more details and signal to noise calculations see the 2dF
WWW pages (http://www.aao.gov.au/cgi-bin/2dfsn.cgi).

\subsection{Special sky subtraction techniques}

While the main effort has gone into optimising the use of 2dF for the
large redshift surveys, i.e. for very large numbers of low dispersion
observations of relatively bright targets, it is clear that 2dF can be
used in many other ways.  Some of the scientific programs are
described in the following section.  Here we mention some technical
factors relevant to other applications of 2dF.

The biggest questions are how faint can 2dF observations be pushed, and
what is the best way of maximising the signal to noise ratio
in a given observing time?
This comes down to two further questions: what is the best way of
doing sky subtraction, and for how long can observations be continued
and still give a useful gain?

The redshift survey data, on which many of the performance figures
quoted above are based, typically consist of sets of $3 \times 1100$s
integrations on sets of galaxies to $B\sim19.5$.  The mean dark sky
signal is approximately equal to 20 per cent or less of the signal for
most galaxies.  For such observations, the standard procedure is to
allocate a fraction (at least 5 per cent) of the fibres to random sky
positions, derive a mean sky and scale this for each fibre.  The same
technique will work for total integration times of several hours and
should produce similar quality data for targets up to one or two
magnitudes fainter.  Beyond that, the observations will be sky limited
and the gains will go as the square root of the observing time, at
best.

Two alternative techniques are classical beam switching, in which
pairs of fibres are placed equal distances apart and each target is
observed alternately in either fibre, and `nod \& shuffle'
\cite{nodding1,nodding2}
where the image is moved rapidly (`shuffled') between an active and a storage
area on the CCD, exactly in phase with the nodding of the telescope
between targets and an offset sky position.  Experiments with both
methods were done with 2dF in January 2001.  In theory, the `nod \&
shuffle' method should give the best results since the identical
optical path is used for the target and sky observations, and because
the switching is very fast compared with the timescale of night sky
line variations.  Beam switching should be almost as good when time
variations are not critical and has the advantage that the target is
always being observed, whereas `nod \& shuffle' is off-source for half
the time.  Both types of switching are inefficient compared with
conventional observing, in terms of the number of targets observable.

In practice neither mode gave dramatic gains in the magnitude range 
accessible to 2dF although sky subtraction can be improved to below 
1 per cent using these methods. A serious drawback of any target-sky 
switching technique is
that it depends on finding clean sky apertures at the same fixed
distance from every target.  This becomes harder as the number of
targets increases.  Moreover, to avoid damaging the signal for an
object at say magnitude 21 requires having no objects in the sky
aperture which are brighter than magnitude 24 or 25, well below the
detection limit of the Schmidt sky surveys or most wide field CCD
images.

Essentially, it depends on whether the limit to the accuracy of sky
subtraction is set by the photon statistics and faint object
contamination in the sky, or by the systematic errors involved in
transferring the sky from other fibres.  In the case of 2dF, it seems
that the operational complications and target losses inherent in beam
switching mean that it is usually better to try to minimise the
systematic errors and to continue to use the mean sky approach.

A more fruitful line may be to exploit the large numbers of objects
observable with 2dF to create mean spectra, rather than trying to
increase the signal to noise ratio in each individual spectrum.  One approach is to take
large numbers of targets selected to be almost identical from other
evidence (e.g. photometry, or an automatic spectral classification scheme)
and combine their spectra; a second is to subdivide samples according
to one or more strong spectroscopic features, and then to combine the
subsets to look for weaker features which may correlate with the
strong features.  Either way, spectra can be created with signal to noise ratio levels
corresponding to hundreds or even thousands of hours of observation.

\subsection{Configuration efficiency}
\label{configperf}

The current version of the {\sc configure} software, which allocates
fibres to targets, has been optimised for the galaxy redshift survey 
\cite{grs}. This is characterised by having a mean target
density of about 180 per square degree, fairly uniformly distributed
across the sky, so that it was necessary to have an algorithm which could
achieve a very high yield in allocating all 400 fibres when there were
often only about 400 targets in the 2dF field.  The galaxy redshift survey 
is able to allocate fibres to 94 per cent of the original input target 
list \cite{grs}.

Somewhat different constraints arise in other applications, e.g. in
crowded globular star clusters, in deep fields from CCD images, or when
there are many fewer than 400 targets.  {\sc configure} has several
options for different ways of allocating fibres to targets and these can
sometimes give substantially higher yields than running the default
parameters.  For example, in crowded compact fields, it is best to centre
the targets and to limit the pivot angle through which fibres can move.
However, the maximum yields in very compact fields are not high, due to
the size and shape of the 2dF buttons and the requirement to leave safe
clearance between buttons and fibres.  Over 300 fibres can be allocated
to targets within a 1\degr\ diameter field and approximately 100 fibres can be
allocated to objects in a 20 arcmin diameter field, based on observations
of dense globular clusters.

For sparse fields with fewer than 200 targets, it is often more efficient
to restrict the fibres to a single spectrograph, if only for ease of data
reduction.  For covering a wide range of magnitudes, e.g. in Galactic open
star clusters, the targets should be split into subsets covering only 2 or
3 magnitudes each.  Note that the reconfiguration time becomes very short
for small samples of stars, so it is not inefficient to use 2dF for
observations of relatively small numbers of bright stars (50 stars take
about 5 minutes to configure).

Other techniques have also been used with the {\sc CONFIGURE} software to
send the light from blue objects to one spectrograph with a blue grating 
and red objects to the second spectrograph with a red optimised grating.
A good example of this is given by Glazebrook \etal \shortcite{config_spectro}.

2dF is not very efficient for doing short observations of single targets
such as standard stars.  There is a procedure for placing a star down any
desired fibre in an existing configuration, but it is rather slow and
cumbersome, and it depends on blind offsetting from a guide fibre to a
target fibre.  If it is desired to measure many such stars, a better
procedure is to prepare a special configuration with just a few fibres in
a simple pattern at the field centre.  For long observations of single
targets where accurate centering is important, it is necessary to have one
or more bright guide stars in the same way as for a full 2dF
configuration.  This applies especially to faint `targets of opportunity' where
it is essential to supply a target position, sky positions and a nearby
guide star.

A key design feature of 2dF, which has been little exploited up to now, is
the ability to keep track of differential atmospheric refraction by
setting up the same configuration on both field plates, but corrected for
different hour angles.  To observe the same field all night one should
tumble between the two plates every hour or two, depending on telescope
attitude (see Fig. \ref{darfig}).  However, setting up such configurations is not
trivial because there are small but significant geometrical differences
between the two 2dF field plates.  These mean that a configuration which
is valid on one plate will fail on the other because of button and fibre
collisions, and because limits on fibre extensions and pivot angles are
exceeded.  These effects can be largely avoided by setting up the starting
configuration with extra clearance around the buttons and reduced limits
on the pivot angle.  The minimum clearance, set by the 2dF hardware, is
400~\micron; setting this to 800~\micron\ and setting the maximum pivot angle to
10\degr\ instead of the default 14\degr\ largely avoids collisions when
switching field plates or changing the configuration HA.  There may be a
decreased target yield since the fibres cannot be placed so close
together, but in practice such losses are generally only a few per
cent.

\section{2dF science}
\label{science}

\begin{figure}
\centering
\psfig{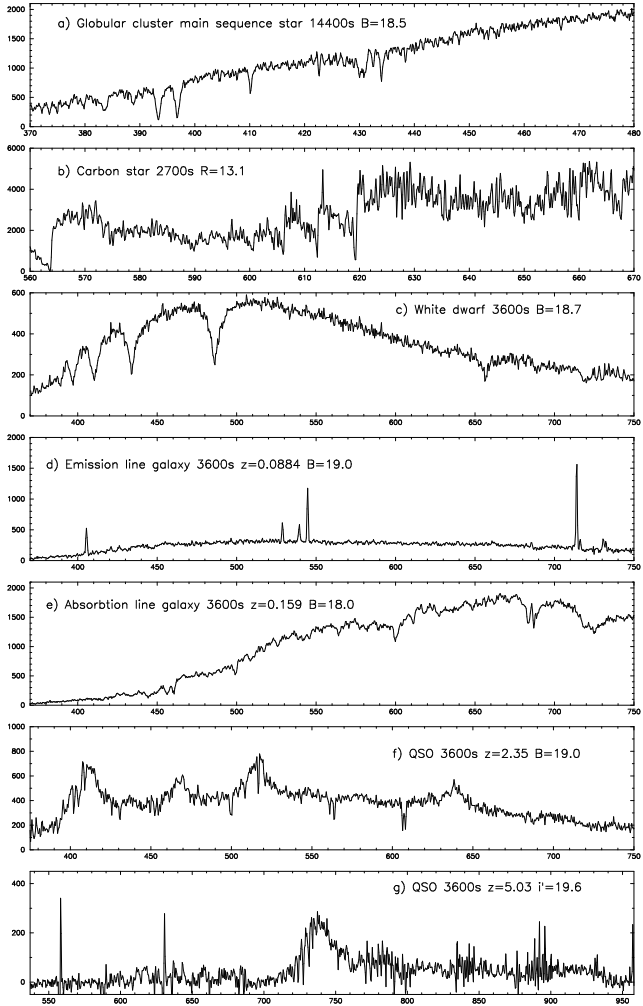}
\caption[dummy]{Some examples of spectra taken with 2dF.  The
first two are high dispersion stellar spectra (1200 l/mm gratings, 2\AA\
resolution), the next four are low dispersion spectra covering
almost the full wavelength range accessible to 2dF, taken with the
300B gratings for the galaxy or quasar surveys.  The final panel 
shows a low dispersion spectrum taken with the 316R grating. The 
exposure times and magnitudes are quoted in each panel. The X axis 
of each panel is the wavelength
in nm, the Y axis is the observed number of counts after data reduction.
 a) Blue spectrum of a faint main
sequence star in the globular cluster 47 Tucanae, in which the
dominant features are the H \& K lines of Ca II near 390nm, the CH
G-band near 430nm and several Balmer H lines; 
b) red spectrum of a
carbon star in the Small Magellanic Cloud, showing the deep Swan C$_2$
band near 560nm and several strong spikes between 605 and 620nm due to
the combined effects of C$_2$ and CN bands (the noisy appearance of this
quite high S/N spectrum is due to a host of real molecular features:
such spectra yield radial velocities with an internal precision of
3~kms$^{-1}$); 
c) a foreground white dwarf star typical of many found in the
colour-selected 2QZ quasar survey, with broad Balmer H absorption
lines; 
d) a strong emission line galaxy from the 2dF Galaxy Redshift Survey, with
lines due to redshifted [OII] near 407nm, H-beta and [OIII] near 540nm
and H-alpha near 710nm; 
e) an absorption line galaxy
with lines due to redshifted  H \& K Ca II absorption near 460nm, 
Mgb absorption at 600nm, Na D near 685nm (next to the
atmospheric B-band O$_2$ absorption feature); 
f) a QSO from the 2QZ survey  with broad
emission lines of Ly$\alpha$ near 410nm, SiIV near 470nm, CIV near 520nm
and CIII] near 640nm. 
g) a QSO identified as part of a joint 2dF--SDSS observing program 
\cite{config_spectro} with broad Ly$\alpha$ near 730nm.
With thanks to S. Croom (private comm.) for the spectra
shown in panels c) and f).
}
\label{spectra}
\end{figure}

While the main science driver for the design of 2dF has been the large redshift
surveys \cite{grs,qrs}, the inherent flexibility in the design
of 2dF means that it is a unique facility for many other types of astronomy.

A diverse range of both galactic and extra-galactic astronomy problems have
already been tackled using the unique area coverage and multiplex
advantage of 2dF. Some of these are described briefly in this section.
Some example spectra obtained with 2dF are shown in Fig. \ref{spectra}.

Gilmore, Wyse, Norris and Freeman are using 2dF to measure radial velocities 
to an accuracy of 10 km$^{-1}$ for several thousand F/G main sequnce 
stars at distances of 3$-$7 kpc from 
the Sun in several sightlines through the Galaxy \cite{aaosps}. 
These data will allow the statistical study of the kinematics
and metallicity distributions of stars in the Galactic thick disk and halo.
Initial results include a new population of low metallicity stars with 
disk-like kinematics, possibly representing debris from a merged satellite.

Spectroscopy of large samples of faint stars in globular clusters have been obtained
by Cannon, Croke, Da Costa and Norris (private comm.).  Their objective is to
compare the chemical abundance variations seen in red giant stars with unevolved
main sequence stars to determine whether these variations are due to
self-enrichment during evolution or are primordial.
In the case of 47 Tucanae,
the spectra show clear separation into CN-strong and CN-weak stars. Combining the
spectra for 50--100 stars in each class yields very high signal to noise ratio
spectra with more than $10^5$ counts per pixel, making it possible to detect
extremely weak features and to look for correlations with other abundance
parameters.

Hatzidimitriou \etal \shortcite{carbonstars} have been using 2dF to study carbon stars in the Magellanic
Clouds. Two observing runs yielded over 2000 carbon star spectra in both
Clouds. These spectra have been used to map in fine detail the  rotation and
velocity dispersion across the LMC. The data are also being used to classify and
determine chemical abundances for the carbon stars.

Evans and Howarth (private comm.) are using 2dF to undertake a spectroscopic survey of massive stars
in the SMC from an unbiased sample of bright blue field stars obtained from APM
photometry. Over 4000 intermediate resolution spectra have already been obtained
and used to generate an observational H-R diagram. This can be compared with
population synthesis models to investigate the field star initial mass function
for the SMC.

Drinkwater \etal \shortcite{fornax} are carrying out a complete, unbiased survey
of the Fornax
cluster by obtaining 2dF spectra for all objects down to B$_j$=19.7 in a 12
square degree region centred on the cluster. The goals of this project are to
determine cluster membership for a complete sample of objects (especially dwarf
galaxies), study the cluster dynamics, detect previously unrecognised compact
galaxies in the cluster and field, and to study background galaxies and quasars
and foreground Galactic stars.

A deep narrowband [OIII] imaging survey of the Virgo cluster has revealed a large
population of emission line objects that could be either planetary nebulae
associated with the intra-cluster medium or high redshift  emission line
galaxies. During a 5 hour 2dF exposure Freeman \etal \shortcite{pne}
obtained 47 emission-line
detections of which 23 turned out to be true intra-cluster planetary nebulae
with detection of both the 4959 and 5007\AA\ lines, and another 16 in
the outer regions of M87.  Freeman \etal also found 8 Ly$\alpha$ emitters at
z$\sim$ 3.1 with equivalent widths W$_\lambda$(Ly$\alpha$) $>$ 150~\AA\ .

Willis \etal \shortcite{willis} have compiled a sample of 485 early-type galaxies with redshifts
with 0.3$\leq$ z $\leq$ 0.6. This represents the largest sample of luminous
field galaxies at intermediate redshift. These data are being used to study the
evolution of galaxy clustering on large scales, for an investigation of the
Fundamental Plane for luminous galaxies as a function of environment, and for a
survey for strong gravitational lensing.

\section{Future plans}
\label{future}

In this paper we have described a unique multi-object optical spectroscopy
facility (2dF) available as a common user instrument at the 
Anglo-Australian Telscope.

We are currently performing experiments using 2dF with techniques such as 
charge shuffling and telescope nodding \cite{nodding2} to improve the 
sky subtraction of extremely  faint target objects.

Future plans include replacing the two spectrographs with bench mounted
spectrographs using volume phase holographic (VPH) grating \cite{vph} 
technology. The VPH gratings have the advantage of offering higher 
efficiency than conventional reflecting diffraction gratings. Upgraded 
detectors, improved anti-reflection  coatings and new optical fibre 
materials \cite{newfibres} will allow us to use longer optical fibres 
and still gain in overall throughput and provide higher resolutions. The
advantage of the bench mounted spectrographs will be in their thermal 
and mechanical stability.

We also plan to use the flexibility of the fieldplate tumbler system to 
provide one or two fibre feeds for integral field spectroscopy at the 
currently unused 90\degr\ positions of the 2dF tumbler unit.

\section{Acknowledgments}

The authors would like to acknowledge the many AAO technical staff who 
have made 2dF a reality and astronomy support staff of the AAO who make 
2dF available to the general user. This project has only been possible 
with extensive support from both the Australian and UK scientific 
communities and the long-term backing of the AAT Board.


\bsp
\label{lastpage}
\end{document}